\documentclass[a4paper, 11pt]{article}

\usepackage{graphicx}
\usepackage{xcolor}
\usepackage{slashed}
\usepackage{amssymb}
\usepackage{mathtools}
\usepackage{latexsym}
\usepackage{textgreek}
\usepackage{verbatim}
 \usepackage[utf8]{inputenc}
 \usepackage{amsmath}
\usepackage{amsfonts}
\usepackage{verbatim}
\usepackage{graphicx}
\usepackage{subfigure}
\usepackage{amssymb}
\usepackage[T1]{fontenc}
\usepackage{slashed}
\usepackage{color}
\usepackage[numbers,sort&compress]{natbib}

\def\beq{\begin{eqnarray}}
\def\eeq{\end{eqnarray}}

\def\({\left(}
\def\){\right)}

\def\mpl{M_{\rm pl}}

\newcommand{\be}{\begin{equation}}
\newcommand{\ee}{\end{equation}}
\newcommand{\la}{\langle}
\newcommand{\ra}{\rangle}
\def\ea{\end{eqnarray}}
\def\ba{\begin{eqnarray}}

\def\beq{\begin{eqnarray}}
\def\eeq{\end{eqnarray}}

\def\({\left(}
\def\){\right)}
\def\mn{_{\mu \nu}}

\def\mpl{M_{\rm pl}}

\def\la{\langle}
\def\ra{\rangle}

\def\lsim{\mathrel{\rlap{\lower3pt\hbox{\hskip0pt$\sim$}}
     \raise1pt\hbox{$<$}}}         
\def\gsim{\mathrel{\rlap{\lower4pt\hbox{\hskip1pt$\sim$}}
     \raise1pt\hbox{$>$}}}

\def\lsim{\mathrel{\rlap{\lower3pt\hbox{\hskip0pt$\sim$}}
     \raise1pt\hbox{$<$}}}         
\def\gsim{\mathrel{\rlap{\lower4pt\hbox{\hskip1pt$\sim$}}
     \raise1pt\hbox{$>$}}}

\begin{document}

\begin{center}
{\Large \bf{Consistent Canonical Quantization of Gravity:\\
Recovery of Classical GR from BRST-invariant Coherent States}}

 \vspace{1truecm}
\thispagestyle{empty} \centerline{\large  {Lasha Berezhiani $^{1,2}$, Gia Dvali $^{1,2}$ and Otari Sakhelashvili $^3$}
}

 \textit{$^1$Max-Planck-Institut f\"ur Physik, F\"ohringer Ring 6, 80805 M\"unchen, Germany\\
 \vskip 5pt
$^2$Arnold Sommerfeld Center, Ludwig-Maximilians-Universit\"at, \\Theresienstra{\ss}e 37, 80333 M\"unchen, Germany\\
 \vskip 5pt
 $^3$Sydney Consortium for Particle Physics and Cosmology, \\
School of Physics, The University of Sydney, NSW 2006, Australia
 }

\end{center}  
 
\begin{abstract}

We perform canonical quantization of General Relativity, as an effective quantum field theory below the Planck scale, within the BRST-invariant framework. We show that the promotion of  constraints to dynamical equations of motion for auxiliary fields leads to the healthy Hamiltonian flow. In particular, we show that the classical properties of Einstein's gravity, such as vanishing Hamiltonian modulo boundary contribution, is realized merely as an expectation value in appropriate physical states. Most importantly, the physicality is shown not to entail trivial time-evolution for correlation functions. In the present approach we quantize the  theory once and for all  around the Minkowaski vacuum and treat other would-be classical  backgrounds as BRST-invariant coherent states. This is especially important for cosmological spacetimes as it uncovers features that are not visible in ordinary semi-classical treatment. The Poincar\'e invariance of the vacuum, essential for our quantization, provides strong motivation for spontaneously-broken supersymmetry. 

\end{abstract}

\newpage
\setcounter{page}{1}

\tableofcontents
\renewcommand{\thefootnote}{\arabic{footnote}}
\setcounter{footnote}{0}

\linespread{1.1}
\parskip 4pt

\section{Introduction}

 This paper is about BRST-invariant formulation of canonically 
 quantized gravity as of consistent low energy effective field theory.  
 The emphasis is made on recovering the essential aspects of classical General Relativity (GR) and in particular the existence of time-evolution, which is sometimes obscured within more ``vintage''  approaches to quantizing GR.  Contrary to semi-classical 
 approaches of quantizing fields on curved classical backgrounds, 
 this approach allows, at least in principle, the resolution of the would-be classical background as of full-fledged quantum state of gravitational degrees of freedom. 

   Namely, the present framework provides an explicit consistent realization  of 
 the previously suggested approach \cite{Dvali:2011aa, Dvali:2012en,Dvali:2013eja,Dvali:2014gua,Berezhiani:2016grw,Dvali:2017eba,Berezhiani:2021zst} implying that the would-be classical metric 
 backgrounds must be treated as the coherent states 
 of gravity quantized on top of the Minkowski vacuum. The analogous treatments of classical backgrounds in 
 quantum field theories of other spins can be found in \cite{Dvali:2013vxa,Dvali:2017ruz,Berezhiani:2020pbv,Berezhiani:2021gph,Dvali:2022vzz,Berezhiani:2023uwt}. 
 
Quantum field theory in curved spacetime \cite{Birrell:1982ix} concerns with quantizing fields, as well as fluctuations of the metric, in classical background geometries. 
Although this approximation is adequate in many cases, it has clear
limitations. In particular, it breaks down over the time scales 
on which the quantum back-reaction on a would-be classical background 
metric becomes significant \cite{Dvali:2011aa,Dvali:2013eja,Dvali:2014gua,Berezhiani:2016grw,Dvali:2017eba, Dvali:2018xpy, Dvali:2018ytn, Dvali:2020wft, Dvali:2024hsb}. 
  As shown in previous papers, this breakdown has important implications for black hole physics and cosmology\footnote{For example, for a recent work on astrophysical implications of so-called black hole ``memory burden effect''  \cite{Dvali:2018xpy, Dvali:2020wft, Dvali:2024hsb}, 
   see \cite{Alexandre:2024nuo,Thoss:2024hsr, Balaji:2024hpu, Haque:2024eyh}.}.     
It has also been shown to have ramifications for the initial conditions for inflation \cite{Berezhiani:2015ola,Berezhiani:2022gnv}.

 The crucial role of analyzing gauge-invariant observables was stressed \cite{Garriga:2007zk} already in connection with the earlier attempts \cite{Tsamis:1996qq,Tsamis:1996qm,Woodard:2004ut,Tsamis:2007is} of accounting for quantum back-reaction 
on de Sitter and inflationary spacetimes. 
In particular, one has to make sure that departures from de Sitter geometry do not represent the artefacts of the choice of gauge for quantum fluctuations.

In this work, we stress the importance of describing the quantum state of the system in its entirety. That is, the background 
spacetime itself is viewed as the quantum state. 
Therefore, we need to quantize gravity before positing the nontrivial background. By quantization we refer to the identification of canonical phase-space degrees of freedom, the symplectic structure, and realization of them as an operator algebra. Upon the discussion of the spectrum of the Hamiltonian, there must indubitably be a special state that would serve as the vacuum of the theory. As we shall argue, the singled-out consistent candidate for such a state is the Minkowski spacetime. 

This way of reasoning selects a certain gauge from the beginning, in which the canonical quantization was performed. Moreover, it sets the gauges of the background and perturbations in a connected way. There is residual freedom to readjust the gauge for perturbations as needed, e.g. in the path-integral, but only for the computation of gauge-invariant quantities. This stems from general principles we are accustomed to in gauge theories, see e.g. \cite{Weinberg}.

This raises important questions about the possibility of switching between frames. Within BRST quantization, quantum states corresponding to the gauge field configuration generically change in such a way that gives vanishing contribution to $S$-matrix elements between physical states.

The paper is organized as follows. In Sec. \ref{backghround_ind}, we kick-start the discussion with a brief outline of the role of having a notion of a well-defined vacuum (like Minkowski) for a consistent canonical quantization of GR, emphasizing that this does not preclude the existence of other consistent states corresponding to nontrivial spacetimes. In Sec. \ref{Smatrix}, we follow this up with a qualitative argument connecting the existence of the low-energy $S$-matrix for multi-graviton states with the existence of the BRST-invariant canonical Hamiltonian formalism. Sec. \ref{SGamma} recounts the perturbative recovery of classical dynamics from the connection between $S$-matrix and an effective action. Sec. \ref{q_GR} presents main results of this work, performing canonical (BRST-invariant) quantization of GR, analyzing the equations of motion and discussing the recovery of the classical dynamics in appropriate quantum states. Sec. \ref{coord_rep} contains the discussion of coordinate reparametrization within the adopted framework. Sec. \ref{cutoff} touches upon the cutoff sensitivity of low energy observables. We conclude in Sec. \ref{conclusion} and discuss the outlook.

\section{On background Independence of Quantization}
\label{backghround_ind}

  Here, we would like to set the framework and clarify the role 
 of the background in our quantization. 
 We shall quantize gravity in theory of GR with zero cosmological term. 
 This does not imply that we are limited by  
 Minkowski as the only possible background.  
 In contrary, once quantized on Minkowski vacuum the theory  
has a full quantum power for accounting for all other consistent backgrounds, including the cosmological ones. 

For example, a homogeneous inflationary background is 
viewed as a coherent state of a scalar field and gravity 
constructed on top of the Minkowski vacuum \cite{Dvali:2013eja, 
Dvali:2014gua, Berezhiani:2016grw, Dvali:2017eba, Berezhiani:2021zst}.

Temporarily, the potential energy of the slow-rolling scalar field 
can be approximated by a positive cosmological term and correspondingly 
the  coherent state 
can be approximated by a pure gravitational de Sitter. The BRST-invariant construction 
of such a state can be found in \cite{Berezhiani:2021zst}. 

  Of course, in practice, it may be easier to study quantum fluctuations 
  around the inflationary state within a conventional semi-classical treatment in which the de Sitter coherent state is treated as a classical 
  background and perturbations are quantized on top of it. 
   Our quantization methods directly apply to these modes.
   
In summary, the quantization of the theory with zero cosmological constant 
does not impair its ability of full quantum description 
of other consistent backgrounds.

   However, there exist independent selection criteria 
 for rejecting certain backgrounds. These are not directly 
 related to our quantization methods, but rather to other aspects 
 of a consistent formulation 
 of the theory. For instance, an eternal de Sitter state does not allow 
 for the  definition of the $S$-matrix. 
 Correspondingly, such vacua are excluded by formulations of the theory 
 based on the $S$-matrix (for the discussion, see \cite{Dvali:2020etd}).  This can act as a powerful selection tool 
 for the parameters of the theory (for various implications, 
 see \cite{Dvali:2024dlb}), but are not 
 essential for the present discussion.

\section{From $S$-matrix to Canonical Hamiltonian Formalism}
\label{Smatrix}

The ultimate goal of this manuscript is the clarification of certain perceived puzzles in canonical quantization of GR as a low-energy effective field theory.

The starting point of the discussion will be the assumption that there exists an $S$-matrix formulation of gravity. This in turn posits the existence of the scattering theory of arbitrary number of gravitons, which are quantum gravitational degrees of freedom defined around the Minkowski spacetime and are represented by the massless spin-2 field.

In order to establish that the aforementioned ultimate $S$-matrix formulation has some semblance of GR at low energies, one would need to match the scattering $S$-matrix obtained from the quantization of the latter within its regime of applicability to the appropriate limit of the former. For this, one should be able to perform a quantization of GR and define the notion of quantum graviton to begin with.

Luckily, such a quantization had been performed and the scattering matrix elements of low energy gravitons had been obtained \cite{Veltman:1975vx}. The starting point is the definition of degrees of freedom. The Poincar\'e symmetry being the guiding principle, the gravitational degrees of freedom are identified as its massless spin-2 representation. This statement, not being the subject of a debates, will be taken for granted. The next, equally important step is the identification of the interaction vertices. It is also a matter of consensus that the self-consistent interactions of said quanta are uniquely fixed to the leading order in derivative expansion, matching the weak-field expansion of GR around the Minkowski spacetime $g\mn=\eta\mn$. Which is the only Poincar\'e invariant classical solution to GR.

In other words, the low-energy scattering amplitudes are obtained from an appropriate quantization of the nonlinear theory of gravitons with the classical action
\beq
S_{\rm EH}=\int d^4x \mathcal{L}_{EH}=\frac12\mpl^2\int d^4 x\sqrt{-g}R(g\mn)\,,
\eeq
with $g\mn\equiv\eta\mn+\mpl^{-1} h\mn$ serving as the definition of the graviton field $h\mn$. This is the classical formulation of gravity, with the coordinate reparameterization invariance corresponding to the gauge redundancy of the formalism.

There are different ways one could proceed quantizing this theory based on the treatment of this gauge redundancy. Modern, consistent, procedures reduce to the BRST-invariant path-integral formulation of scattering amplitudes in the Lagrangian formalism. The generating functional can be schematically given as
\beq
Z[T^{\mu\nu},\ldots]=\int\left[\mathcal{D}h\right] \left[\mathcal{D}c\right]\left[\mathcal{D}\bar{c}\right]\left[\mathcal{D}b\right]e^{-i\int d^4 x \left(\mathcal{L}_{\rm EH}+\mathcal{L}_{\rm GF}+\mathcal{L}_{\rm FP}+h\mn T^{\mu\nu}+\ldots\right)}\,,
\label{ZT}
\eeq
where $\mathcal{L}_{\rm GF}$ and $\mathcal{L}_{\rm FP}$ stand for the gauge-fixing and ghost Lagrangians respectively. As it is customary, the auxiliary field $b_\mu$ imposing the gauge-fixing condition and the Hermitian Faddeev-Popov ghosts ($c^\mu$ and $\bar{c}_\mu$) have been introduced. In Eq. \eqref{ZT}, $T^{\mu\nu}$ is the external current for $h\mn$ and ellipsis stand for the external currents for the aforementioned auxiliary sector. In this section, we keep the discussion schematic without being specific about the detailed form of various terms, which can be found in section \ref{q_GR}. This schematic expression is also obscuring the contribution connected to the noncanonical nature of the kinetic term of GR, which is immaterial for tree-level computations and for the point we are trying to make here.

The generating functional \eqref{ZT} can be, and has been, used to compute the in-out correlation functions and scattering amplitudes for low-energy gravitons. Having the low-energy $S$-matrix from the aforementioned Lagrangian path-integral formalism is very good, but in order to connect it with real-time physical observables we need to identify the Hamiltonian. The usual quantization procedure begins with the latter and upon integrating out conjugate momenta results in the former. However, since the canonical quantization raises more eyebrows these days than the Lagrangian path-integral formalism \eqref{ZT}, we can retrace our steps backwards. From \eqref{ZT}, we proceed by integrating in the canonical conjugate momenta for the spatial metric degrees of freedom $h_{ij}$ and ghosts. We also need to keep in mind that $b_\mu$ is related to the canonical conjugate of $h_{0\mu}$ via a linear transformation, as we will see explicitly in section \ref{canonical_variables}, that has a unit Jacobian. Not surprisingly, following this procedure we recover the classical Hamiltonian of GR upon reintroducing the missing conjugate momenta by means of the Gaussian integral, which is due to the fact that the Hamiltonian is at most quadratic in canonical momenta (as it is well known and reiterated in section \ref{canonical_variables}). The resulting schematic form of the generating functional is
\beq
Z[J^{\mu\nu},\ldots]=\int\left[\mathcal{D}h\right]\left[\mathcal{D}\Pi\right] \left[\mathcal{D}c\right]\left[\mathcal{D}\Pi_c\right]\left[\mathcal{D}\bar{c}\right]\left[\mathcal{D}\Pi_{\bar{c}}\right]e^{-i\int d^4 x \left( \dot{h}_{\mu\nu}\Pi^{\mu\nu}+\dot{c}^\mu \Pi^c_\mu+\dot{\bar{c}}_\mu\Pi_{\bar{c}}^\mu-H+h\mn J^{\mu\nu}+\ldots\right)}\,,
\eeq

After all is said and done, we end up with the path-integral in phase-phase where all degrees of freedom have the canonical conjugate momenta. We can use the result to identify the Hamiltonian of the system in the field and conjugate momentum eigenstate bases, which we can then use to build the Hamiltonian operator. This should be highly indicative that the canonical quantization framework in phase-space should have merit and therefore be able to account for the time-dependence of physical quantities. In other words, the existence of the low-energy scattering $S$-matrix of gravitons is in one-to-one correspondence with the existence of the well-defined Hamiltonian quantization within appropriate regime of validity. 

We would like underline that not all canonical quantization attempts are equal. What we have outlined in this section is how to recover BRST-invariant canonical formalism of \cite{Kugo:1978rj} from the Lagrangian path-integral \eqref{ZT}. This should not come as a surprise as the BRST symmetry is organic to the latter as well. However, some of the canonical quantization attempts entail such a peculiar treatment of the constraints of the GR that they would be impossible to recover from \eqref{ZT}.

\section{Recovery of Classical GR from $S$-matrix}
\label{SGamma}

 In this section we recount the connection between the $S$-matrix 
 and effective action. In particular, the classical effective action 
 in the background gravitational field created by a set of classical sources 
 is given by the expectation value of the $S$-matrix operators
 over coherent states describing these  sources. Schematically
 this is given by the general equation: 
              \begin{equation}  \label{StoS}
\la coh | \hat{S} | coh \ra = \Gamma_{eff} \,. 
 \end{equation} 
   where $\hat{S}$ is the total $S$-matrix operator implying summation 
   over all $1/\mpl$ contributions and $| coh \ra$ stands 
   for corresponding coherent states. 
    Since we are mainly interested in the recovery 
    of classical gravity from quantum theory of graviton,  
    we shall present the derivation in $1/\mpl$ expansion 
    taking into account the contributions into $\hat{S}$ coming 
    from tree-level graviton exchanges. 
   The corresponding discussions can be found, e.g., in 
   lectures \cite{Dvali:lecture}.

Let us start with linearzed Einstein gravity. The
Lagrangian is, 
     \begin{eqnarray}  \label{Leff}
\mathcal{L} \, = \, \frac{1}{2}  h^{\mu\nu} {\mathcal E}_{\mu\nu} \,  - \,  \frac{1}{\mpl} \, h_{\mu\nu}  T^{\mu\nu} \,,
   \end{eqnarray}    
where, 
     \begin{eqnarray}  \label{hEq}
 {\mathcal E}_{\mu\nu} &\equiv & \, \Box h_{\mu\nu} - \eta_{\mu\nu} h -
  \partial_{\mu}\partial^{\beta} h_{\nu\beta}  
 -   \partial_{\nu}\partial^{\beta} h_{\mu\beta}  \nonumber \\  
  && + 
   \eta_{\mu\nu} \partial^{\alpha}\partial^{\beta} h_{\alpha\beta}   +  \partial_{\mu}\partial_{\nu} h \,, 
\end{eqnarray}    
is the linearized Einstein tensor and 
$T_{\mu\nu}$ is a classical energy momentum source. 
In quantum theory, this source should be understood 
as the expectation value of an operator 
$\hat{T}_{\mu\nu}$ over a certain coherent state, 
$| coh \ra$, such that  $\la coh | \hat{T}_{\mu\nu}  | coh \ra  = 
    T_{\mu\nu}$.
The operator $\hat{T}_{\mu\nu}$ is composed 
of field operators that source gravity. 

 As it is well-known, the above Lagrangian describes a unique linear
 ghost-free theory of a massless spin-$2$ field. 
 It matches the theory obtained ``top-down'' by linearization of 
 the Einstein-Hilbert action.  In this sense, the Einstein theory 
 passes an immediate test of uniqueness and of quantum consistency at the linear level.

 The corresponding equations of motion have the form, 
    \begin{eqnarray}  \label{EQH}
 {\mathcal E}_{\mu\nu}\, = \, \frac{1}{\mpl} \,T_{\mu\nu} \,.
 \end{eqnarray}
In de Donder gauge ($\partial^{\mu} h_{\mu\nu} = \frac{1}{2}
 \partial_{\nu} h$) this equation 
  takes the form: 
     \begin{eqnarray}  \label{DSLin}
\Box \, h_{\mu\nu} \,  = \,  \frac{1}{\mpl} (T_{\mu\nu} -  
 \frac{1}{2}\eta_{\mu\nu} T) \,,    
 \end{eqnarray} 
 and is solved by 
       \begin{eqnarray}  \label{LinSol}
 h_{\mu\nu}^{(1)}(x)  \,  = \, \frac{1}{\mpl}\,  \int d^4 x_1 \,  \Delta_{\mu\nu,\alpha\beta}(x-x_1) \, T^{\alpha\beta} (x_1) \,,    
 \end{eqnarray} 
where   
      \begin{eqnarray}  \label{propagator1} 
  \Delta_{\mu\nu,\alpha\beta} (x-x_1) = 
  \int \frac{d^4p}{(2\pi)^4} \, {\rm e}^{-ip(x-x_1)} \, 
  \frac{\frac{1}{2}(\eta_{\mu\alpha} \eta_{\nu\beta}  + 
  \eta_{\mu\beta}  \eta_{\nu\alpha}) - 
 \frac{1}{2} \eta_{\mu\nu} \eta_{\alpha \beta}}{p^2 } 
   \end{eqnarray}
is the Green's function evaluated with retarded pole-prescription.  The superscript on $h\mn$ indicates the 
order in $1/\mpl$. 
 
In order to monitor the gravitational field, we can introduce 
a probe energy-momentum tensor denoted by  $\tau_{\mu\nu}$. 
  For simplicity we can assume that $\tau_{\mu\nu}$ is arbitrarily soft, so that the 
higher order effects in it can be safely ignored. However, this is not essential, since these effects can be consistently taken into account whenever needed. 
 In classical theory $\tau_{\mu\nu}$ is composed out 
 of some classical fields. 

To the linear order in $\tau_{\mu\nu}$, 
its effective action in the classical gravitational field is given by 
        \begin{eqnarray}  \label{StauT}
A_{eff} \, = && \, \frac{1}{\mpl} \int d^4x\, h_{\mu\nu}(x) \,\tau^{\mu\nu}(x)  \,.    
 \end{eqnarray} 
After taking the variational derivative of the full action with respect to 
degrees of freedom composing $\tau_{\mu\nu}$, we obtain 
their equation of motion in the background gravitational field. 

 This procedure is trivially lifted to a quantum theory
 in which the fields composing the probe are promoted into quantum operators and so is their energy-momentum tensor  
 $\tau_{\mu\nu} \rightarrow \hat{\tau}_{\mu\nu}$.
  Correspondingly, the action (\ref{StauT}) is promoted into 
  a quantum effective action in a background classical field 
         \begin{eqnarray}  \label{StauQ}
A_{eff}(q) \, = && \, \frac{1}{\mpl} \int d^4x\, h_{\mu\nu}(x) \,\hat{\tau}^{\mu\nu}(x)  \,.    
 \end{eqnarray}  
 
The actions (\ref{StauT}) and (\ref{StauQ}) can be reconstructed 
perturbatively as series in $1/\mpl$.
  To the order $1/\mpl^2$, the action of the classical probe $\tau_{\mu\nu}$ moving in the classical metric  $h_{\mu\nu}(x)$ created by the source $T_{\mu\nu}$, is 
        \begin{eqnarray}  \label{Stau}
A_{eff}^{(2)} \, = && \, \frac{1}{\mpl} \int d^4x\,  \,h_{\mu\nu}^{(1)}(x) \,\tau^{\mu\nu}(x)  \,   \nonumber \\
= &&\, \frac{1}{\mpl^2} \,  \int \, d^4 x\, dx_1^4 \,  \tau^{\mu\nu}(x)  \,
 \Delta^{\rm F}_{\mu\nu,\alpha\beta}(x-x_1) \,  T^{\alpha\beta}(x_1)  \,,   
 \end{eqnarray} 
where 
      \begin{eqnarray}  \label{propagator2} 
  \Delta^{\rm F}_{\mu\nu,\alpha\beta} (x-x_1) = 
  \int \frac{d^4p}{(2\pi)^4} \, {\rm e}^{-ip(x-x_1)} \, 
  \frac{\frac{1}{2}(\eta_{\mu\alpha} \eta_{\nu\beta}  + 
  \eta_{\mu\beta}  \eta_{\nu\alpha}) - 
 \frac{1}{2} \eta_{\mu\nu} \eta_{\alpha \beta}}{p^2 \, + \, i\epsilon} 
   \end{eqnarray}
   is the Feynman propagator.

   This expression is equal to the expectation value of 
   the quantum $\hat{S}$-matrix operator of one-graviton exchange 
           \begin{eqnarray}  \label{S2}
\hat{S}^{(2)} \, = \, \frac{1}{\mpl^2} \int d^4x d^4x_1 \,  \hat{\tau}^{\mu\nu} \langle \hat{h}_{\mu\nu}(x)
\hat{h}_{\alpha\beta}(x_1) \rangle \,  \hat{T}^{\alpha\beta}(x_1)\,,    
 \end{eqnarray}   
   taken over the coherent 
    state $|coh\ra = |coh\ra_{\tau} \times |coh\ra_{T} $    
    of quantum degrees of freedom composing 
    the operators  $\hat{\tau}_{\mu\nu}$ and $\hat{T}_{\mu\nu}$. 
    That is, the coherent states satisfy, 
    $\la coh | \hat{\tau}_{\mu\nu}  | coh \ra  = 
    \tau_{\mu\nu}$ and $\la coh | \hat{T}_{\mu\nu}  | coh \ra  = 
    T_{\mu\nu}$. For simplicity, we assume that the 
    sources $\hat{\tau}_{\mu\nu}$ and $\hat{T}_{\mu\nu}$ are composed of independent degrees of freedom, so that 
    the coherent state factorizes into two independent ones. 
    However, this is not essential.  
   In the above expression the graviton operators are fully contracted giving the Feynman propagator, 
            \begin{eqnarray}  \label{Fprop}
 \langle \hat{h}_{\mu\nu}(x)
\hat{h}_{\alpha\beta}(x_1) \rangle \, 
= \,  \Delta^{\rm F}_{\mu\nu,\alpha\beta}(x-x_1)\,.   
 \end{eqnarray}

    It is then obvious that we have the relation, 
            \begin{equation} 
\la coh | \hat{S}^{(2)}  | coh \ra = A_{eff}^{(2)} 
 \end{equation} 
  In other words, the quantum $S$-matrix operator generates 
  the classical gravitational action for the sources. 
  
   This connection holds to all orders in $1/\mpl$ expansion \cite{Deffayet:2001uk}. 
   That is, the contribution of the higher order non-linearities 
   to the classical metric are in one to one correspondence 
   with quantum $S$-matrix
   elements generated by tree-graviton exchanges among the 
   classicalized  sources to the same order in $1/\mpl$. 
   
  For example,   consider an effect of an addition of a 
  cubic-order graviton vertex to the Lagrangian, which we shall schematically denote by $\frac{1}{\mpl}(\partial^2 h(x)h(x)h(x)) $, accounting for the presence of two derivatives and $1/\mpl$. The tensorial structure which is rather lengthy is not shown explicitly. 
  
  Variation of this vertex with respect to $h_{\mu\nu}$ 
  corrects the equation (\ref{EQH}) by a bilinear term, 
      \begin{eqnarray}  \label{EQH2}
 {\mathcal E}_{\mu\nu}\, = \, \frac{1}{\mpl} \,T_{\mu\nu}
 +   \frac{1}{\mpl} (\partial^2 h^2)_{\mu\nu} \,.
 \end{eqnarray}  
   The leading order effect of this term is an order $1/\mpl^3$ correction to $h_{\mu\nu}(x)$ which can be taken into account 
iteratively,  and gives, 
       \begin{eqnarray}  \label{Sol2}
 h_{\mu\nu}^{(3)} (x) \,  = \, \frac{1}{\mpl}\,  \int d^4 x_1 \,  \Delta_{\mu\nu,\alpha\beta}(x-x_1) \, (\partial^2 (h^{(1)}(x_1))^2)^{\alpha\beta} \,,    
 \end{eqnarray} 
where, the $h^{(1)}(x_1)$ is a solution (\ref{LinSol}). 
Clearly,  (\ref{Sol2}) is of order $1/\mpl^3$ and the corresponding correction to the effective action 
        \begin{eqnarray} 
A_{eff}^{(4)} \, = \, \frac{1}{\mpl} \, \int d^4x\,  \,h_{\mu\nu}^{(3)}(x) \,\tau^{\mu\nu}(x)  \,,    
 \end{eqnarray} 
is of order $1/\mpl^4$. 
 Again, this effective action is matched by 
 the expectation value of the 
 $\hat{S}$-matrix operator to the same order, 
            \begin{eqnarray}  
\hat{S}^{(4)} \, = \, \sum \, \frac{1}{\mpl^4}\, \int d^4x d^4x_1d^4x_2 d^4x_3 \,  \hat{\tau}^{\mu\nu} \hat{h}_{\mu\nu}(x)
\hat{h}_{\alpha\beta}(x_1) \hat{T}^{\alpha\beta}(x_1)
\hat{h}_{\gamma\delta}(x_2) \hat{T}^{\gamma\delta}(x_2)
(\partial^2 (h(x_3)^3)
\,,    
 \end{eqnarray}   
 where summation is taken over all possible 
 combinations in which  all graviton operators are 
 singly-contracted.   The expectation  value is taken 
 over the same coherent states that classicalize the sources. 
 This gives  
              \begin{equation}  \label{Stau4}
\la coh | \hat{S}^{(4)}  | coh \ra = A_{eff}^{(4)} \,.
 \end{equation} 

\begin{figure}
 	\begin{center}
\includegraphics[height=0.2\textwidth]{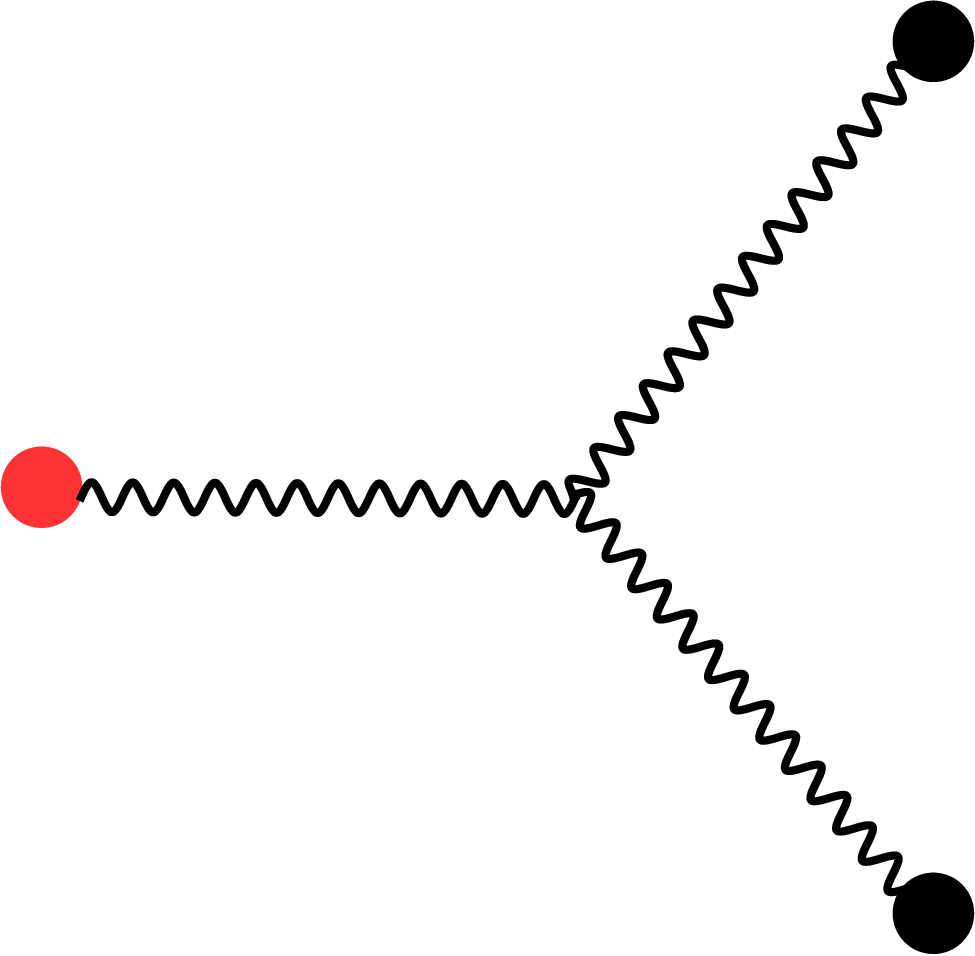}\hskip 30pt \includegraphics[height=0.2\textwidth]{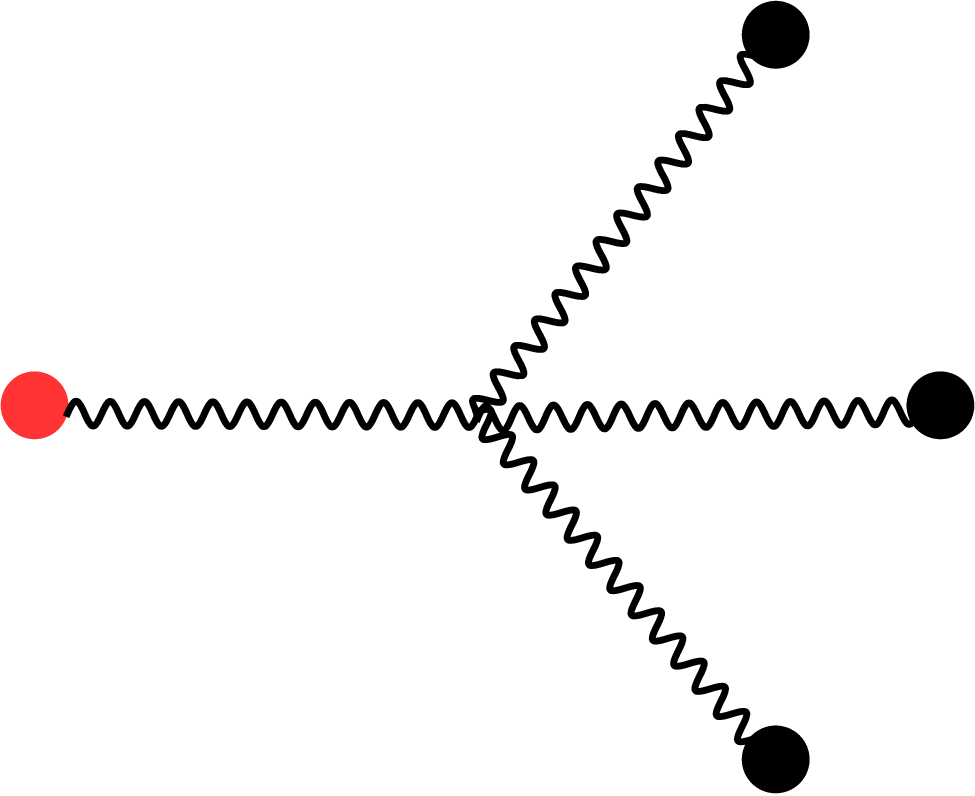}\hskip 30pt \includegraphics[height=0.2\textwidth]{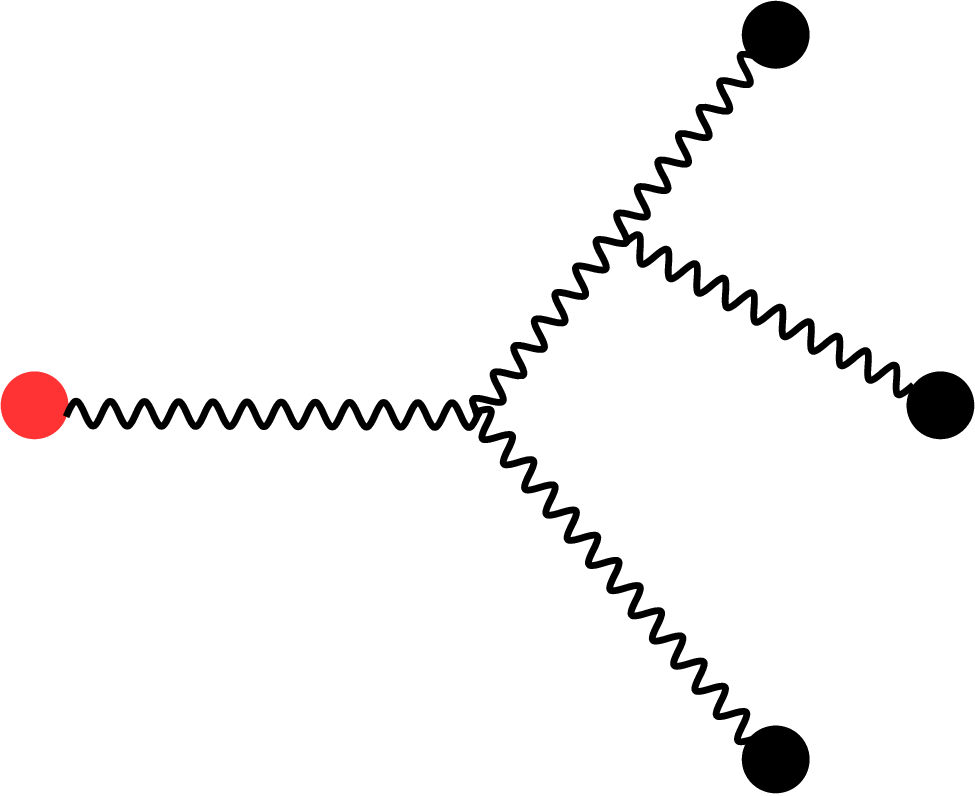}
 		\caption{ A diagrammatic examples of order
		$1/\mpl^4$ and $1/\mpl^6$ 
  contributions 
		to the effective action of the probe $\tau_{\mu\nu}$ denoted by the red dot.  The source $T_{\mu\nu}$ is denoted by the black dots.  } 	
\label{processes}
 	\end{center}
 \end{figure}

 Of course, already at the level of a cubic interactions 
 we get contributions at arbitrary higher orders in $1/\mpl$. 
  The example of a $1/\mpl^4$ and $1/\mpl^6$ contributions are given in Fig. \ref{processes}. 
 
 With increasing order of non-linearities in the Lagrangian 
 the number of contributions is increasing drastically but 
order by order the same relation holds.  
This is understandable, since with gravitons appearing 
as virtual states among the classical sources, the classical 
contribution in each order in $1/\mpl$ must be accounted by the tree-level exchanges.

 For example, denoting  a $k$-graviton interaction vertex
at spacetime point $x$  by  $V(h(x)^k)$
 let us consider  a generic $\hat{S}$-operator which is 
 order $n$ in $1/\mpl$,  first order in the probe $\tau_{\mu\nu}$ 
 and $l$-th order in the source $T_{\mu\nu}$ and 
  with all sources interacting 
 with gravitons linearly.  
 This operator has a form, 
             \begin{eqnarray}  
\hat{S}^{(n)} \, =  &&\,  \frac{1}{\mpl^n} \sum \int d^4x \, d^4x_1\, ... \,d^4x_l \,  
d^4y_1\,...\, d^4y_q \nonumber \\
 &&\, \hat{\tau}^{\mu\nu}(x) \hat{h}_{\mu\nu}(x) \, 
\hat{T}^{\alpha\beta}(x_1) \hat{h}_{\alpha\beta}(x_1) \, ... \, 
\hat{T}^{\gamma\delta}(x_l) \hat{h}_{\gamma\delta}(x_l) \, 
\nonumber \\ 
&& V(h(y_1)^{k_1}) \, ... \,  V(h(y_q)^{k_q}) 
\,,    
 \end{eqnarray}   
 The summation goes over all possible contractions as well as  
 over various insertions of vertices that satisfy
 $n = l + 1 + \sum_{i=1}^q (k_i -2)$.  Again, all gravitons are 
 singly-contracted.  
  After classicalizing the sources by taking the expectation values 
  over proper coherent states, this expression maps on a corresponding effective action of the probe. 
    The mapping is straightforwardly generalized to the case of 
    nonlinear interactions between the sources and gravitons
    as long as all the graviton operators are singly-contracted.

 We thus arrive to a general schematic relation, 
             \begin{equation}  \label{StauN}
\la coh | \hat{S}^{(n)}  | coh \ra = A_{eff}^{(n)} \,,
 \end{equation} 
that holds in $n$-th order in $1/\mpl$ expansion.

Strictly speaking, the connection must hold beyond weak field regime. 
The only caveat is that beyond weak-field the $1/\mpl$-expansion breaks down and the series must be re-summed.  
 
      The above result is a particular manifestation of 
      the generic relation between the effective action and 
      the $S$-matrix which holds for an arbitrary quantum field 
      $\hat{\phi}$.

\section{Quantization of GR}
\label{q_GR}

To proceed, we need to choose the framework. We employ the BRST-invariant formulation of GR, contrary to a more commonly thought after Wheeler-DeWitt quantization\footnote{Some potential differences between  Wheeler-DeWitt and BRST quantizations have been discussed in the literature (see, e.g., \cite{VanHolten:2001nj,Witten:2022xxp}).  The full scope of 
the distinction between the two frameworks shall become transparent 
from the analysis below.}. Its initial development goes back to \cite{Kugo:1978rj,Kugo:1979gm}, where the recovery of Einstein's equations together with various other aspects were discussed.

In this framework, the gauge-fixed Lagrangian density takes the following form
\beq
&&\mathcal{L}=\mathcal{L}_{\rm EH}+\mathcal{L}_{\rm GF}+\mathcal{L}_{\rm FP}\,,\\
&&\mathcal{L}_{\rm EH}=\sqrt{-g} \mpl^2R\,,\\
\label{gauge_fixing}
&&\mathcal{L}_{\rm GF}=\mpl b_\nu\partial_\mu\left(\sqrt{-g} g^{\mu\nu}\right) - \frac{1}{2}\alpha\eta^{\mu\nu}b_\mu b_\nu\,,\\
\label{ghostlagrangian}
&&\mathcal{L}_{\rm FP}=i \partial_\mu \bar{c}_\nu \left(\sqrt{-g}g^{\mu \sigma}\partial_\sigma c^{\nu}-\partial_\sigma\left(\sqrt{-g}g^{\sigma\nu}\right) c^{\mu}\right)\,.
\eeq
Here $\eta_{\mu\nu}$ is the Minkowski metric and $\alpha$ is a free parameter which plays the same role as $\xi$ in $R_\xi$-gauge of quantum electrodynamics. In fact, we could have simply chosen $\alpha=0$. As it is customary, we have employed the auxiliary field $b_\mu$ to impose the de Donder gauge and have introduced the Hermitian ghost fields $c^\mu$ and $\bar{c}_\mu$ accordingly.

The reason the introduction of gauge-fixing auxiliary fields facilitates the canonical quantization is because in the gauge-invariant formulation the temporal components of the metric lack the conjugate momenta.  In the presence of $b_\mu$, on the other hand, the temporal metric components acquire canonical counterparts.

In the Lagrangian formalism it is convenient to work with $\tilde{g}^{\mu\nu}\equiv\sqrt{-g}g^{\mu\nu}$, as it was done in  \cite{Kugo:1978rj}. In terms of this variable, the Einstein-Hilbert Lagrangian can be written in the following convenient form
\beq
\mathcal{L}_{\rm EH}=\mpl^2\left(\tilde{g}^{\rho\sigma}\tilde{g}_{\lambda\mu}\tilde{g}_{\kappa\nu}-2\delta_\kappa^\sigma\delta^\rho_\lambda\tilde{g}_{\mu\nu}-\frac{1}{2}\tilde{g}^{\rho\sigma}\tilde{g}_{\mu\kappa} \tilde{g}_{\lambda\nu} \right)\partial_\rho \tilde{g}^{\mu\kappa}\partial_\sigma \tilde{g}^{\lambda\nu}\nonumber\\
+2\mpl^2\partial_\mu\left( \frac{1}{2}\tilde{g}^{\mu\nu}\tilde{g}_{\alpha\beta}\partial_\nu \tilde{g}^{\alpha\beta}+\partial_\nu\tilde{g}^{\mu\nu} \right)\,.
\eeq
However, it is easy to see that the theory has been written in such a way that there are time derivatives of the temporal components of the metric even in the absence of the gauge-fixing sector. This makes the formulation of the canonical formalism somewhat cumbersome. Therefore, in order to find the more convenient parametrization for the temporal components, we switch to ADM formalism.

\subsection{ADM Formalism}
\label{ADM}

The variables of ADM decomposition \cite{Arnowitt:1959ah} are lapse $N$, shift $N_j$ and the spatial metric $\gamma_{ij}$, which are introduced as follows
\beq
g\mn=
\begin{pmatrix}
-N^2+N_kN^k & N_j \\
N_i & \gamma_{ij}
\end{pmatrix}\,,\qquad
g^{\mu\nu}=
\begin{pmatrix}
-N^{-2} & N^{-2} N^j \\
N^{-2} N^i & \gamma^{ij}-N^{-2}N^iN^j
\end{pmatrix}\,,
\eeq
(with $\gamma_{ik}\gamma^{kj}=\delta_i^j$ and $N^i=\gamma^{ij}N_j$). In these variables the Einstein-Hilbert part reduces to
\beq
&&\mathcal{L}_{\rm EH}=\sqrt{-g} \mpl^2R\nonumber\\
&&~~~~~~=\mpl^2\left(N\sqrt{\gamma}\left( K_{ij}K^{ij}-K^2+^{(3)}R\right)-2\partial_t\left(\sqrt{\gamma} K\right)+2\partial_i(\sqrt{\gamma}(KN^i-\gamma^{ij}\partial_jN))\right)\,,
\eeq
where
\beq
K_{ij}=\frac{1}{2N}\left(\nabla_i N_j+\nabla_j N_i-\partial_t \gamma_{ij} \right)\,.
\eeq
As it is customary, the total derivatives have been isolated in such a way that $N$ and $N_j$ are free of derivatives, everywhere except on the boundary. After the adjustment of the boundary term following the Gibbons--Hawking--York procedure, the only surviving boundary contribution is precisely what doctor prescribed to recover the ADM mass for the asymptotically Minkowski geometry. As it plays no role for the bulk dynamics, we will ignore it for the time being and reintroduce it when relevant
\beq
\mathcal{L}_{\rm EH}=\mpl^2N\sqrt{\gamma}\left( K_{ij}K^{ij}-K^2+^{(3)}R\right)\,.
\eeq
Next, we proceed with the identification of the convenient canonical degrees of freedom.

\subsection{Canonical Variables}
\label{canonical_variables}

For convenience, we choose the canonical fields to be $\gamma_{ij}$ and $A^{\mu}\equiv\sqrt{-g}g^{0\mu}$. The latter are related to ADM variables by
\beq
N=-\frac{\sqrt{\gamma}}{A^0}\,,\qquad {\rm and} \qquad N^j=-\frac{A^j}{A^0}\,.
\eeq
This choice follows from the examination of the Lagrangian for ghosts \eqref{ghostlagrangian}. The classical expression for the conjugate momentum of $\gamma_{ij}$, in the absence of the ghost sector, is given by
\beq
\label{piij}
\Pi^{ij}=-\mpl^2\sqrt{\gamma}\left( K^{ij}-\gamma^{ij}K\right)\,.
\eeq
One of the advantages of rewriting the ghost Lagrangian of \cite{Kugo:1978rj} into \eqref{ghostlagrangian} by integration by parts is to maintain this classical relation. Moreover,  had we chosen $N$ and $N_j$ as canonical variables instead of $A^\mu$,  we would have gotten an additional contribution to \eqref{piij} from the ghost sector.
 The conjugate momentum of $A^\mu$ is readily given by
\beq
\Pi_\nu=\mpl b_\nu-i(\partial_\mu \bar{c}_\nu ) c^\mu\,.
\label{pinu}
\eeq
As for the ghosts, we have
\beq
\label{pic}
&&\Pi_{\bar{c}}^\nu=i\left( A^\sigma \partial_\sigma c^\nu-\partial_\sigma(\sqrt{-g}g^{\sigma\nu})c^0 \right)\,,\\
&&\Pi^c_\nu=-iA^\mu\partial_\mu\bar{c}_\nu\,,
\label{picbar}
\eeq
where we defined the conjugate momenta using the left-differentiation. This way, the Hamiltonian is obtained as
\beq
H=\int d^3 x\left( \dot{A}^\mu \Pi_\mu+\dot{c}^\mu \Pi^c_\mu+\dot{\bar{c}}_\mu\Pi_{\bar{c}}^\mu+\dot{\gamma}_{ij}\Pi^{ij}-\mathcal{L}\right)\,.
\eeq
It must be noted that the time-derivative of temporal components is undetermined from the canonical variables, due to the fact that they only enter in $\Pi_{\bar{c}}^\nu$ together with $\dot{c}$. However, as it is customary in such cases, the undetermined time-derivative drops out from the Hamiltonian. In particular, upon collecting relevant terms
\beq
H\supset \int d^3x \dot{A}^\nu\left( \Pi_\nu-\mpl b_\nu+i\partial_\mu \bar{c}_\nu c^\mu \right)\,,
\eeq
which vanishes in the light of \eqref{pinu}.

Although $\dot{A}^\nu$ drops out from the Hamiltonian it is nevertheless determined from Hamilton's equation for $A^\mu$, in a complete analogy with the Coulomb potential of electrodynamics. It is worthwhile to mention that the relevant equation follows from the gauge-fixing terms of the Lagrangian. 

Moving forward, upon simplification the ghost and gauge-fixing contributions to the Hamiltonian take the following form
\beq
\label{HFPGF}
H_{\rm FP+GF}=\int d^3 x \Big[ \frac{i}{A^0}\left\{\Pi^c_\nu+iA^j\partial_j\bar{c}_\nu\right\}\left\{\Pi_{\bar{c}}^\nu-iA^k\partial_k c^\nu\right\} -i \partial_i\bar{c}_\nu \sqrt{-g}g^{ij} \partial_j c^\nu \nonumber\\
-\Pi_0 \partial_k A^k-\Pi_i \partial_j\left( \sqrt{-g}g^{ij} \right)+\frac{1}{2}\alpha\eta^{\mu\nu} b_\mu b_\nu\Big]\,,
\eeq
where we keep in mind that $\sqrt{-g}g^{ij}=\frac{1}{A^0}\left(- \gamma \gamma^{ij}+A^iA^j\right)$. Also, we have kept the last term written in terms of $b_\nu$, despite the fact that it is not a canonical degree of freedom, for notational convenience. Furthermore, this is an expendable term as we can always focus on $\alpha=0$ case.

The Einstein-Hilbert part of the Hamiltonian is well known and up to the boundary term reduces to
\beq
\label{H_EH}
H_{\rm EH}=\int d^3x \left[-\frac{1}{A^0} \mathcal{H}+\frac{A^i}{A^0}\mathcal{P}_i\right]\,,
\eeq
where $\mathcal{H}$ and $\mathcal{P}^j$ stand for the so-called Hamiltonian and momentum constraints respectively
\beq
\label{hamiltonianconst}
&&\mathcal{H}\equiv\frac{1}{2\mpl^2}\left( \gamma_{ik}\gamma_{j\ell}+\gamma_{i\ell}\gamma_{jk}- \gamma_{ij}\gamma_{k\ell}\right)\Pi^{ij}\Pi^{kl}-\mpl^2\gamma R^{(3)}\,,\\
&&\mathcal{P}_i\equiv-2\gamma_{ik}\partial_j\Pi^{kj}-(2 \partial_k\gamma_{ji}-\partial_i\gamma_{jk})\Pi^{jk}\,.
\label{momentumconst}
\eeq

Once again, we would like to emphasize that the boundary term, responsible for reproducing the ADM mass for the asymptotically flat geometry, must be added; upon relevance, we will comment about its necessity.
So far the discussion has been classical, that is why we have not yet payed attention to the order of bosonic operators in the Hamiltonian. The next step is to promote fields to operators, assigning the canonical (anti-)commutation relations and to define the vacuum. It is worth emphasizing that the latter condition will be indicative to how the field operators defined so far differ from the operators corresponding to the elementary quantum degrees of freedom. This difference will be merely a constant shift, which does not affect the canonical (anti-)commutation relations. These points will be further clarified as we proceed with quantization.

\subsection{Canonical Quantization}

Within the adopted framework, all components of the metric possess nonzero canonical conjugates. Therefore, we proceed in a usual fashion by assigning the following equal-time (anti-)commutation relations
\beq
&&\left[\hat{\gamma}_{ij}(x),\hat{\Pi}^{k\ell}(y)\right]=\frac{i}{2}\left(\delta_{i}^{k}\delta_{j}^{\ell}+\delta_{i}^{\ell}\delta_{j}^{k}\right)\delta^{(3)}(x-y)\,,\\
&&\left[\hat{A}^\mu(x),\hat{\Pi}_\nu(y)\right]=i\delta^\mu_\nu\delta^{(3)}(x-y)\,,\\
&&\left\{\hat{c}^\mu(x),\hat{\Pi}^c_\nu(y)\right\}=i \delta^\mu_\nu\delta^{(3)}(x-y)\,,\\
&&\left\{\hat{\bar{c}}_\nu(x),\hat{\Pi}_{\bar{c}}^\mu(y)\right\}=i \delta^\mu_\nu\delta^{(3)}(x-y)\,.
\eeq
Here hats indicate the promotion of fields to operators, while $[\ldots]$ and $\{\ldots\}$ indicate the commutation and anti-commutation respectively.
Notice that unlike QED and the linearized gravity, $b_\mu$ is not strictly speaking the conjugate momentum of the temporal degrees of freedom $A^\mu$ due to the additional ghost contribution to \eqref{pinu}. However, it is determined in terms of $\Pi_\mu$ and $\Pi^c_\mu$, using \eqref{picbar}. 

Let us emphasize that when promoting the Hamiltonian to the operator
\beq
\label{totalH}
\hat{H}=\hat{H}_{\rm EH}+\hat{H}_{\rm FP+GF}\,,
\eeq
we need to be cautious with ordering non-commuting fields. For starters it needs to be Hermitian, a task which is rather straightforward to achieve.  
We would like to note that non-commuting operators enter the Hamiltonian at most in quartic order. In fact, the first term of \eqref{hamiltonianconst} is the only source of the term containing four non-commuting quantities. Notice that this is the virtue of our approach to quantize $A^\mu$ rather than the corresponding ADM variables $N$ and $N_j$. In particular, using the shift vector would introduce the inverse metric in terms that involve the conjugate momentum of the spacial metric. For the rest of this work we will not pay close attention to the ordering of noncommuting operators at all times, as it is not relevant for our discussion. We will be implicitly implying that the corresponding terms have been ordered appropriately to ensure Hermiticity.

This brings us to the next important point, concerning the vacuum of the theory. If the BRST quantization procedure we have adopted has any merit to it, then the lowest energy eigenstate $|\Omega\ra$ of the Hamiltonian (which incorporates the boundary contribution resulting in the ADM mass) must correspond to the Minkowski spacetime\footnote{In the classical limit this is achieved via positivity theorem \cite{Witten:1981mf}. For the full quantum theory, the same can be achieved using Supergravity, which in turn is motivated by the vacuum structure of GR \cite{Dvali:2024dlb}.}, i.e.
\beq
\label{gammaexp}
&&\la \Omega | \hat{\gamma}_{ij} |\Omega \ra=\delta_{ij}\,,\\
\label{a0exp}
&&\la \Omega | \hat{A}^0 |\Omega \ra=-1\,,\\
\label{ajexp}
&&\la \Omega | \hat{A}^j |\Omega \ra=0\,,\\
&&\la \Omega | \hat{b}_\mu |\Omega \ra=0\,.
\label{bexp}
\eeq
It must be stressed that the last condition must hold in any physical state, not merely in the vacuum. The reason for this is the same as in QED, it follows from the BRST transformation properties of the anti-ghost field. Namely, we have
\beq
\label{bQc}
\hat{b}_\nu=i\{\hat{Q},\hat{\bar{c}}_\nu\}\,,
\eeq
implying $\la\psi|\hat{b}_\nu|\psi\ra=0$ in any state satisfying $\hat{Q}|\psi\ra=0$. Here, $\hat{Q}$ is the Noether charge of the BRST symmetry, and physical Hilbert space is defined as a set of all zero-BRST charge eigenstates with vanishing ghost number. As it straightforwardly follows from \eqref{gammaexp} and \eqref{a0exp}, quantum fields corresponding to the elementary excitations around the vacuum are defined as
\beq
\label{elementmetric}
&&\hat{h}_{ij}\equiv \hat{\gamma}_{ij}-\delta_{ij}\,,\\
&&\hat{a}^0\equiv \hat{A}^0+1\,.
\label{elementtemp}
\eeq
In other words, the creation-annihilation operators creating particles in the vacuum should be defined by the decomposition of $\hat{h}_{ij}$, $\hat{a}^0$ and $\hat{A}^j$ (together with ghost fields). Therefore we could have simply started with these shifted fields when introducing equal-time commutation relations. Obviously this would not have changed anything, as both set of fields satisfy the same commutation relations.

It is straightforward to demonstrate that the canonical formulation at hand leads to the quantum Einstein's equations for the operators supplemented with the gauge-fixing and ghost terms. In order to show without a doubt that the classical dynamics can be recovered from these equations, one needs to construct BRST-invariant states corresponding to classical geometries, with the prime candidates being coherent states. 

 The explicit construction of such states is challenging, however we can analyze the expectation value of the above mentioned equation of motion in physical state to see if the physicality of the state automatically entails the inconsistency with classical dynamics.

In fact we will argue (in Sec. \ref{rep_class}) that the BRST-construction 
recovers classical  nonlinear equations of motion for 1-point functions supplemented with additional quantum terms that can be treated perturbatively in $\hbar$ for certain backgrounds.

Before we delve into discussion of how the classical dynamics is recovered within the adopted framework, it is imperative to discuss the question of the gauge choice. For this, we only need the part of the Hamiltonian. This is an important point for understanding the introduction of classical backgrounds in different coordinates.

\subsection{Gauge Freedom}

The relevant equation follows from the gauge-fixing sector and in the Lagrangian formalism descends from varying the action with respect to $b_\nu$. In the canonical operator framework, on the other hand, it comes from Hamilton's equation for temporal degrees of freedom
\beq
\label{Adot}
\dot{\hat{A}}^\mu=i[\hat{H},\hat{A}^\mu]\,.
\eeq
This is both the gauge-fixing equation and the equation defining $\dot{\hat{A}}^\mu$. Luckily, we have the time-derivative of only the anti-ghost field in \eqref{pinu}. As such, we can straightforwardly deduce
\beq
\left[\hat{A}^\mu(x),\hat{\Pi}_\nu(y)\right]=i\delta^\mu_\nu\delta^{(3)}(x-y)\,,\qquad \Longleftrightarrow \qquad \left[\hat{A}^\mu(x),\mpl \hat{b}_\nu(y)\right]=i\delta^\mu_\nu\delta^{(3)}(x-y)\,.
\eeq
Rewriting the relevant part of the Hamiltonian in terms of $\Pi_\nu$, we have
\beq
\hat{H}\supset \int d^3 x \left( - \hat{\Pi}_0 \partial_j\hat{A}^j- \hat{\Pi}_i\partial_j\left(\frac{\hat{\gamma} \hat{\gamma}^{ij}}{\hat{A}^0}\right)-\frac{1}{2\hat{A}^0}\hat{A}^i(\partial_i \hat{\Pi}_j+\partial_j \hat{\Pi}_i)\hat{A}^j+\frac{1}{2}\alpha\eta^{\mu\nu} \hat{b}_\mu \hat{b}_\nu \right)\,.
\eeq
Here we have ordered the third term to ensure the hermiticity. Also, $1/\hat{A}^0$ needs to be understood as a Taylor series expansion in $\hat{a}^0\equiv 1+\hat{A}^0$.

As a result \eqref{Adot} takes the following form in three-dimensional decomposition
\beq
&&\dot{\hat{A}}^0=-\partial_j \hat{A}^j+\frac{\alpha}{\mpl} \hat{b}_0\,,\\
&&\dot{\hat{A}}^i=\partial_j\left(\frac{1}{\hat{A}^0}\left[\hat{\gamma} \hat{\gamma}^{ij}-\hat{A}^i\hat{A}^j\right]\right)-\frac{\alpha}{\mpl}\hat{b}_i\,.
\eeq
As we have already stated,  $\hat{b}_\mu$ has vanishing matrix element between physical states. This property is a direct consequence of \eqref{bQc}. We are interested in the expectation value of the metric since it serves as a classical proxy for the quantum state. Therefore, we have
\beq
\label{A0exp}
&&\partial_t\la\psi |\hat{A}^0|\psi\ra=-\partial_j \la\psi |\hat{A}^j|\psi\ra\,,\\
&&\partial_t\la\psi |\hat{A}^i|\psi\ra=\partial_j\la\psi |\left(\frac{1}{\hat{A}^0}\left[\hat{\gamma} \hat{\gamma}^{ij}-\hat{A}^i\hat{A}^j\right]\right)|\psi\ra\,,
\label{Ajexp}
\eeq
where $|\psi\ra$ is a physical state satisfying $\hat{Q}|\psi\ra=0$. As we can see, \eqref{A0exp} is very similar to the QED gauge-fixing condition. Linearity simplifies the story since it means that the 1-point expectation value satisfies the classical constraint. The other three gauge-fixing conditions \eqref{Ajexp}  are not as pure and it seems that for certain states may include quantum corrections. However, the absence of quantum corrections may be possible to demonstrate using perturbative expansion in $\hbar$, utilizing the background field method and realizing that the correlation functions at coincidence should have vanishing gradients. Although this is not guaranteed if the state in question corresponds to an inhomogeneous background. This is an interesting point, the investigation of which we postpone to future work.

Here, we would like to focus on states corresponding to classical configurations, i.e. states with macroscopic occupancy with nonvanishing 1-point function in $\hbar\rightarrow 0$ limit. For such states, the leading and most relevant part of the gauge-fixing condition \eqref{Ajexp} reduces to
\beq
\partial_t\la\hat{A}^i\ra=\partial_j\left(\frac{1}{\la\hat{A}^0\ra}\left[\la\hat{\gamma}\ra \la\hat{\gamma}^{ij}\ra-\la\hat{A}^i\ra \la\hat{A}^j\ra\right]\right)+\mathcal{O}(\hbar)\,.
\label{Ajconst}
\eeq
It is important to keep in mind that for well-defined physical states
one can choose the coordinate frames which satisfy $\la \hat{A}^0 \ra<0$.

As expected \eqref{A0exp} and \eqref{Ajconst} are equivalent to
\beq
\partial_\mu \left(\sqrt{-g_{\rm cl}}g_{\rm cl}^{\mu\nu} \right)=0\,,
\label{dedondercl}
\eeq
at the classical level, where $g^{\rm cl}_{\mu\nu}$ is the 1-point expectation value of the metric. This equation could have been derived bypassing the Hamiltonian formalism, as in \cite{Kugo:1978rj}, by promoting the Euler-Lagrange equation for $b_\mu$ to an operator equation and evaluating the expectation value in the BRST-invariant state.

\subsection{Reproducing Classical Dynamics}
\label{rep_class}

The above naturally brings us to the discussion of how BRST formalism recovers the classical dynamics, which is an important consistency check of the approach. 
Let us begin with the aspects of GR that are traditionally considered to cause issues upon quantization. The relevant equations consist of Hamiltonian and momentum constraints
\beq
\label{clHconst}
&&\mathcal{H}=0\,\\
\label{clPconst}
&&\mathcal{P}_i=0\,,
\eeq
explicitly given by \eqref{hamiltonianconst} and \eqref{momentumconst} respectively, together with dynamical equations for the spacial metric and its conjugate momentum. Interestingly, the satisfaction of constraints entails vanishing of the Einstein-Hilbert Hamiltonian density away from the boundary, given by the integrand of \eqref{H_EH}. Notably, the theory admits time-dependent configurations that satisfy constraints and have vanishing bulk Hamiltonian density. In classical theory there is no contradiction in this observation.

The question arising in quantum theory concerns the implementation of constraints. Namely, if one chooses to define the Hilbert space of physical states by
\beq
\label{qH}
&&\mathcal{\hat{H}}|phys\ra=0\,\\
&&\mathcal{\hat{P}}_i|phys\ra=0\,,
\label{qP}
\eeq
then they are annihilated by $\hat{H}_{\rm EH}$ of \eqref{H_EH}, in the absence of the boundary term. As a result, one is led to the conclusion that the subsequent quantum gravity is incapable of generating a nontrivial Hamiltonian flow in physical Hilbert space \cite{DeWitt:1967yk}. Consequently, the quantization at hand reproduces one of the properties of its classical counterpart, by yielding vanishing bulk contribution to the Hamiltonian from physical states. 
 
 However, even upon the introduction of the appropriate boundary terms, the main challenge is the reproduction of the time-dependent expectation values of metric degrees of freedom (away from the boundary) in physical states. 
 
A successful quantization must reproduce both of the above-mentioned properties of the classical theory. Namely, there must exist a notion of vanishing Hamiltonian (up to the boundary contribution) in the classical limit, while permitting a time-dependent one-point function of the metric degrees of freedom. The only possibility for achieving these within BRST formalism is by satisfying the following:
\begin{itemize}

\item $\lim_{\hbar\rightarrow 0}\la phys| \hat{H} | phys\ra=0$, where $\hat{H}$ is given by \eqref{totalH} and lacks the boundary term that yields the ADM mass. In order for \eqref{totalH} to satisfy this condition, we need to show that the auxiliary sector gives a vanishing contribution in the classical limit. 

\item $\hat{H}|phys\ra\neq 0$, for the states with non-trivial bulk dynamics.

\end{itemize}

Now, the preceding argument should not be perceived as the statement about the irrelevance of the aforementioned boundary condition that yields the ADM mass. In fact, this contribution becomes irreplaceable for describing the configurations that are eternal in the classical limit.  A good example of such a configuration is a black hole.  
Classically, a black hole is an eternal state in the sense that its characteristics, such as the mass or the angular momentum,  do not change in time. 
  However, in quantum theory it evolves via Hawking evaporation
and this evolution inevitably triggers a departure from the classical description.  There exist excellent arguments indicating that 
after a certain critical time, referred to as ``quantum break-time'', the departure becomes order-one, invalidating the semi-classical picture \cite{Dvali:2013eja, Dvali:2018xpy, Dvali:2020wft}. 
 However, the magnitude of the deviation is not central for the 
 present  discussion.  The key point is that, no matter what, the difference 
 between classical and quantum evolution is non-zero. 
 Correspondingly, the question arises how the ADM mass-fixing boundary Hamiltonian accounts for it. 

 Within the presented picture of BRST quantization, the 
 above connection is guaranteed from the fundamental principles of formulation.   
  That is, the would-be classical dynamics  
 is recovered by the time-evolution of a proper 
 coherent state describing the object. 
 At the same time, the account for the departure 
 of the true quantum evolution from the classical one
 is accounted by an inevitable loss of coherence. 

 For instance, a classical gravitational wave which can be 
 initially approximated by a coherent state of gravitons, 
 shall depart from classical evolution due to a quantum re-scattering 
 of its constituents. Of course, the corresponding 
 quantum break-time can be extremely long, but this does not change the essence of the issue.  

 In other words, in the BRST approach, the main consistency test 
 is the recovery of classical approximation within a certain time scale, as the existence of quantum corrections to such evolution 
 are built-in by the very framework.

Let us now demonstrate the recovery of the classical dynamics from 
BRST quantization more explicitly.
   Within this framework, the fate of constraints stems from Hamilton's equations for the canonical conjugate to the temporal degrees of freedom
\beq
\partial_t \hat{\Pi}_\mu=i[\hat{H},\hat{\Pi}_\mu]\,.
\eeq
The resulting expression for the time and space components of this equation take the following form
\beq
\label{Hconst}
&&\partial_t \hat{\Pi}_0=-\frac{1}{(\hat{A}^0)^2}\hat{\mathcal{H}}+\frac{\hat{A}^i}{(\hat{A}^0)^2}\hat{\mathcal{P}}_i+\frac{1}{(\hat{A}^0)^2}\partial_j \hat{\Pi}_i\left(-\gamma \gamma^{ij}+\hat{A}^i\hat{A}^j\right)\nonumber\\
&&~~~~~~~~~~+\frac{i}{(\hat{A}^0)^2}\left[ \hat{\Pi}^c_\nu \hat{\Pi}^\nu_{\bar{c}}+i\hat{A}^k\left(\partial_k\hat{\bar{c}}_\nu \cdot\hat{\Pi}^\nu_{\bar{c}}-\hat{\Pi}^c_\nu\cdot\partial_k\hat{c}^\nu \right)+\gamma \gamma^{ij}\partial_i\hat{\bar{c}}_\nu\cdot \partial_j\hat{c}^\nu\right]\,,\\
\label{Pconst}
&&\partial_t \hat{\Pi}_j=-\frac{1}{\hat{A}^0}\hat{\mathcal{P}}_j-\partial_j \hat{\Pi}_0-\frac{1}{\hat{A}^0}\left[ \left(\partial_j\hat{\Pi}_k+\partial_k \hat{\Pi}_j\right)\hat{A}^k+\hat{A}^k\left(\partial_j\hat{\Pi}_k+\partial_k \hat{\Pi}_j\right) \right]\nonumber\\
&&~~~~~~~~~~+\frac{1}{\hat{A}^0} \left( \partial_j\hat{\bar{c}}_\nu\cdot\Pi^\nu_c-\hat{\Pi}^c_\nu\cdot \partial_j \hat{c}^\nu \right)\,.
\eeq
In the current operatorial form these expressions represent the dynamical equations rather than constraints. The reason is that
 within the presented 
quantization scheme all the metric components are treated 
as the dynamical degrees of freedom. However, as it is well known,
upon acting on physical states, some of the equations generate constraints. In fact, similar to the recovery of Gauss' constraint in QED, the quantum generalization of Hamiltonian and momentum constraints are recovered as matrix elements of the aforementioned equations in between the physical states.

In order to extract the classical Hamiltonian and the momentum constraints as the leading effects, we need to make certain assumptions about physical states. This would not be necessary if we were able to perform an explicit construction of BRST-invariant (non-asymptotic) states. 

We begin by noticing that, since one-point functions for ghost fields should vanish in physical states, upon bracketing \eqref{Hconst} and \eqref{Pconst} with such states, the leading order ghost contribution should be in the form of their two-point functions at coincidence. It is important to keep in mind that some of such corrections will vanish for kinematic reasons while the others will be divergent and would have to be absorbed in counter-terms, with a possible finite leftovers. Either way, all such contributions must be dropped in the classical limit.
 
In order to simplify the rest of the terms, we would like to use properties of $\hat{b}_\mu$ utilizing \eqref{pinu} with the promotion to operators with proper ordering
\beq
\hat{\Pi}_\nu=\mpl \hat{b}_\nu+\frac{1}{2\hat{A}^0}[\hat{\Pi}^c_\nu, \hat{c}^0]+i\frac{\hat{A}^j}{\hat{A}^0}\partial_j\hat{\bar{c}}_\nu\cdot \hat{c}^0-i\partial_j\hat{\bar{c}}_\nu \cdot\hat{c}^j\,.
\eeq
Correspondingly, in the classical limit, the ghost contributions 
must be dropped from the expectation value of this quantity. 
 If we further employ the fact that $\la phys|\hat{b}_\mu|phys \ra=0$, we find the vanishing expectation value for $\hat{\Pi}_\mu$. Therefore, the expectation values of \eqref{Hconst} and \eqref{Pconst} reduce to
\beq
&&-\left\la\frac{1}{(\hat{A}^0)^2}\hat{\mathcal{H}}\right\ra+\left\la\frac{\hat{A}^i}{(\hat{A}^0)^2}\hat{\mathcal{P}}_i\right\ra+\left\la\frac{1}{(\hat{A}^0)^2}\partial_j \hat{\Pi}_i\left(-\gamma \gamma^{ij}+\hat{A}^i\hat{A}^j\right)\right\ra+\mathcal{O}(\hbar)=0\\
&&\left\la\frac{1}{\hat{A}^0}\hat{\mathcal{P}}_j\right\ra+\left\la\frac{1}{\hat{A}^0}\left[ \left(\partial_j\hat{\Pi}_k+\partial_k \hat{\Pi}_j\right)\hat{A}^k+\hat{A}^k\left(\partial_j\hat{\Pi}_k+\partial_k \hat{\Pi}_j\right) \right]\right\ra+\mathcal{O}(\hbar)=0\,.
\eeq
Moreover, since we are interested in states that correspond to classical configurations built over the Minkowski vacuum, we make further assumptions about the correlation functions of canonical degrees of freedom $\{\hat{h}_{ij},\hat{\Pi}_{ij},\hat{a}^0,\hat{A}^j,\hat{\Pi}_\mu\}$ (defined around \eqref{elementmetric} and \eqref{elementtemp}) that describe the excitations on top of the vacuum. Namely, we posit that the quantum state in question corresponds to the state of the system with a non-vanishing one-point function for these operators. At the same time, their higher order correlation functions are assumed to be  dominated by disconnected contributions. This assumption can be justified perturbatively in $\mpl^{-1}$, following the construction of \cite{Berezhiani:2021zst}. In other words, we are assuming that there exist BRST-invariant coherent states for which the 
composite operators satisfy 
\beq
\left\la f\left(\hat{h},\hat{\Pi},\ldots\right)\right\ra=f\left(\la\hat{h}\ra,\la\hat{\Pi}\ra, \ldots\right) \left( 1+\mathcal{O}(\hbar) \right)\,.
\eeq
Then, the equations further reduce to
\beq
\label{semiclassH}
&&-\frac{1}{\la\hat{A}^0\ra ^2}\mathcal{H}\left(\la \hat{\gamma}_{k\ell} \ra,\la \hat{\Pi}_{mn} \ra\right)+\frac{\la\hat{A}\ra ^i}{\la\hat{A}^0\ra ^2}\mathcal{P}_i\left(\la \hat{\gamma}_{k\ell} \ra,\la \hat{\Pi}_{mn} \ra\right)+\mathcal{O}(\hbar)=0\\
\label{semiclassP}
&&\frac{1}{\la\hat{A}^0\ra}{\mathcal{P}}_j\left(\la \hat{\gamma}_{k\ell} \ra,\la \hat{\Pi}_{mn} \ra\right)+\mathcal{O}(\hbar)=0\,.
\eeq
Under these assumptions, one can easily convince oneself that the expectation value of the Hamiltonian \eqref{totalH} vanishes in the classical limit. In other words, we have the expression
\beq
\la \hat{H} \ra=\mpl^2\int d^3x \Big[-\frac{1}{\la \hat{A}^0\ra} \mathcal{H}\left(\la \hat{\gamma}_{k\ell} \ra,\la \hat{\Pi}_{mn} \ra\right)\{1+\mathcal{O}(\hbar)\}+\frac{\la\hat{A}^i\ra}{\la\hat{A}^0\ra}\mathcal{P}_i\left( \la \hat{\gamma}_{k\ell} \ra,\la \hat{\Pi}_{mn} \ra \right)\{1+\mathcal{O}(\hbar)\}\nonumber \\
+\mathcal{O}(\hbar)\Big]\,,
\eeq
which vanishes in $\hbar\rightarrow 0$ limit, in light of \eqref{semiclassH} and \eqref{semiclassP}.

Having demonstrated the recovery of one of the classical properties of GR, it is important to appreciate that this does not directly lead to
\beq
\hat{H}|phys\ra \stackrel{?}{=} 0\,.
\eeq
Obviously, this condition would be much stronger than what we have demonstrated for the expectation value. Within the adopted framework the constraints \eqref{qH} and \eqref{qP} are replaced by a single constraint based on the BRST charge
\beq
\label{brstconst}
\hat{Q}|phys\ra=0\,.
\eeq
Now, it is important to acknowledge that this constraint does bare a resemblance to \eqref{qH} and \eqref{qP}, just as its QED counterpart contains the operator of Gauss' law within it, see e.g. \cite{Berezhiani:2021zst}. In particular, even though
\beq
\hat{Q}_{\rm QED}=\int d^3 x \left[\hat{c}\left( g\hat{\rho}-\partial_j \hat{E}_j\right)+\hat{B}\hat{\Pi}_{\bar{c}}+\partial_j\left( \hat{c}\hat{E_j} \right)\right]\,,
\eeq
the constraint \eqref{brstconst} does not entail
\beq
\left( g\hat{\rho}-\partial_j \hat{E}_j\right)|phys\ra\stackrel{?}{=}0\,,\label{WD-el_mg_pa}
\eeq
because of the auxiliary sector. In fact, 
for the BRST quantization, the constraint equation \eqref{WD-el_mg_pa} takes the following form
\begin{equation}
    \left( g\hat{\rho}-\partial_j \hat{E}_j\right)|phys\ra=-i \{ \hat Q_{QED},\hat\Pi_c\}|phys\ra\,,
\end{equation}
and 
differs from \eqref{WD-el_mg_pa} due to 
the nontrivial contribution from the right-hand side. The outlined difference is crucial, as it tells us that the BRST quantization has a larger Hilbert space than a naive constraint quantization.

The Gauss law is recovered between arbitrary physical states,
\begin{equation}
  \la phys'|\left( g\hat{\rho}-\partial_j \hat{E}_j\right)|phys\ra=-i\la phys'| \{ \hat Q_{QED},\hat\Pi_c\}|phys\ra\,=0\,\label{Supercanonical_BRST}
\end{equation}
even for different $|phys\ra$ and $|phys'\ra$.
This equation shows the difference between BRST and the aforementioned naive quantization. The latter puts all the constraint on the state appearing from one side, where in the BRST quantization it is ``shared'' between ket- and bra-states. 

This feature is well-known to be exhibited in the Gupta-Bleuler quantization. The gauge-fixing condition is set on the Hilbert space partially. The positive-frequency part is set on the ket-states, while the negative-frequency part is imposed on the bra-states. This is naturally realized in the BRST quantization, since the right-hand side of \eqref{Supercanonical_BRST} can be related to the gauge condition using
\begin{equation}
    i\{ \hat Q_{QED},\hat\Pi_c\}=\dot{\hat{B}},
\end{equation}
where $\hat{B}$ is equal to the gauge-fixing condition based on Hamilton's equation, and it vanishes only between physical states.

Let us reiterate that, unlike the naive constraint quantization, in the BRST framework Gauss' law is realized dynamically. The constraint appears from a dynamical equation of motion, and is fulfilled only between two physical states. All of the above suggests that gravitational Hamiltonian $\hat{H}$ does not necessarily annihilate the physical states and is capable of generating the time evolution.

The dynamics follows from spatial Einstein's equations which take the following form 
\beq
\partial_t \hat{\Pi}^{mn}=\frac{2}{\mpl^2}\frac{1}{\hat{A}^0}\left( \hat{\Pi}^{mk}\hat{\Pi}_k^n-\frac{1}{2}\hat{\Pi}^{mn}\hat{\Pi} \right)+\partial_k\left(\frac{\hat{A}^m}{\hat{A}^0}\right)\hat{\Pi}^{nk}+\partial_k\left(\frac{\hat{A}^n}{\hat{A}^0}\right)\hat{\Pi}^{mk}- \partial_i\left(\frac{\hat{A}^i}{\hat{A}^0}\hat{\Pi}^{mn}\right)\nonumber\\
-\frac{\mpl^2}{A^0}\hat{\gamma}\left( \hat{\gamma}^{mn} \prescript{{(3)}}{}{\hat{R}}+\prescript{(3)}{}{\hat{R}}^{mn} \right)-\mpl^2\hat{\gamma}\left(\nabla^m\nabla^n-\gamma^{mn}\nabla^2\right)\frac{1}{\hat{A^0}} \nonumber \\
+\frac{1}{2}\frac{\hat{\gamma} }{\hat{A}^0}\hat{\gamma}^{mi}\hat{\gamma}^{nj} \left( \partial_i \hat{\Pi}_j+\partial_j \hat{\Pi}_i-\partial_i\hat{\bar{c}}_\nu\partial_j\hat{c}^\nu-\partial_j\hat{\bar{c}}_\nu\partial_i\hat{c}^\nu \right)\nonumber\\
+\frac{\hat{\gamma}}{\hat{A}^0}\hat{\gamma}^{mn}\left( \partial_k \hat{\Pi}^k-\partial_k\hat{\bar{c}}_\nu\partial^k \hat{c}^\nu \right)\,.
\label{dynamical_Eq}
\eeq
At this point it is straightforward to see that upon computing the expectation value of this equation in a physical state and following the same line of reasoning as in recovering the classical constraint equations, 
we will arrive at the classical equations for $\la\hat{\Pi}^{mn}\ra$ which will coincide with the corresponding Einstein's equations. In particular, for a generic physical state for which the right-hand side \eqref{dynamical_Eq} has a nonzero disconnected contribution at $\hbar^0$-order, we obtain
\beq
\lim_{\hbar\rightarrow 0}\partial_t \la\hat{\Pi}^{mn}\ra\neq 0\,.
\eeq
We have thus demonstrated that BRST quantization of gravity 
correctly reproduces the classical dynamics. Upon reintroduction of quantum terms, the corresponding correction to the dynamics can be straightforwardly obtained within this framework. However, this goes beyond the scope of the current work and will be considered elsewhere.

\section{Reparameterization of Coordinates}
\label{coord_rep}

Let us now discuss the question of reference frames within the canonical quantization of gravity at hand. In particular, any physical state corresponding to a certain classical geometry must satisfy the gauge-fixing condition introduced upon quantization. As a result, the configurations corresponding to different frames are still allowed, however not all classical slices of a geometry are reproducible by the 1-point expectation value in a physical state. In fact, even if we construct a BRST-invariant state for which 1-point function reproduces a classical geometry in a certain frame at a given moment of time, the classical evolution away from the original slicing of this 1-point function will be governed by the gauge-fixing condition \eqref{dedondercl}. It goes without saying that the states corresponding to different frames are physically-equivalent, as it can be demonstrated by the equivalence of the corresponding $S$-matrix elements. Moreover, as it is well known, the correlation functions of gauge-invariant quantities are independent of the gauge-fixing condition. However, we are often interested in non-gauge-invariant correlation functions; especially in cosmology. 
Furthermore, the standard gauge-invariant variables in cosmological perturbation theory are invariant to the linear order.

We begin demonstrating some of these points by considering the vacuum state corresponding to Minkowski geometry. In the following sections, we will also see the implications for FLRW Universe.

\subsection{From Minkowski to Rindler}

The Minkowski spacetime $g_{\mu\nu}=\eta_{\mu\nu}$ obviously satisfies de Donder gauge condition \eqref{dedondercl}. An accelerated observer views the same spacetime in the Rindler coordinates $x^\mu_{\rm r}$, which are related to Minkowski coordinates $x^\mu$ by
\beq
t_{\rm r}=\frac{1}{a}{\rm arctanh}\left(\frac{t}{x}\right)\,,\qquad x_{\rm r}=\sqrt{x^2-t^2}\,,
\eeq
where $a$ stands for the acceleration in the positive $x$ direction and the observer is at rest at $x=1/a$ at $t=0$. In these coordinates the spacetime interval takes the following form
\beq
ds^2=-(a x_{\rm r})^2 dt_{\rm r}^2+dx_{\rm r}^2+dy_{\rm r}^2+dz_{\rm r}^2\,.
\eeq
Due to covariance of Einstein's equations, classically the Rindler metric
\beq
\label{rindler}
g_{\mu\nu}={\rm diag}\left(-(a x)^2,1,1,1\right)\,
\eeq
is as good a solution as the Minkowski space $g_{\mu\nu}=\eta_{\mu\nu}$. However it is straightforward to see that this metric does not satisfy \eqref{dedondercl}. This implies that if we construct a physical coherent state in which the expectation value of the metric at some moment of time $t$ is given by \eqref{rindler},
the equality will not be maintained at later times even in the classical limit.

In order to ensure that $\la \hat{g}_{\mu\nu} \ra$ maintains
a given classical form, 
we need to adjust the gauge-fixing term from the beginning accordingly. For example, had we imposed  $\partial_\mu (g^{\mu\nu})=0$ instead of \eqref{dedondercl}, the Rindler metric  \eqref{rindler} would have been a proper background at all times, at least classically. Incidentally, so would the Minkowski space itself. 

The corresponding modification of the gauge-fixing term \eqref{gauge_fixing} would have been
\beq
\label{cosmogauge}
\mathcal{L}_{\rm GF}'=\mpl b_\nu\partial_\mu g^{\mu\nu} - \frac{1}{2}\alpha\eta^{\mu\nu}b_\mu b_\nu\,,
\eeq
with appropriately adjusted ghost terms. Which in turn would force us to reconsider the parametrization of the temporal degrees of freedom of the metric. As we are about to see, this particular gauge-fixing is relevant for the FLRW background as well.

\subsection{FLRW Cosmology}

Although the explicit construction of fully nonlinear BRST-invariant states is challenging, we can make certain statements about the consistency of the gauge-fixing sector with backgrounds of cosmological significance.

    We shall now discuss implications of our  quantization framework 
  to cosmology.  First, as any other would-be classical state,  
  in this framework a cosmological background represents an
   expectation value of the metric operator over a coherent state. 
   The metric  operator is obtained by BRST quantization 
   of the metric field once and for all on a special state  
   representing the vacuum of the theory (in the present case 
   Minkowski).  The coherent state describing the cosmological background is also constructed on top of this vacuum \cite{Berezhiani:2021zst}.

    In this language, the entire cosmology is captured by the 
    quantum evolution of the state.  This in particular encodes  
  the information about all possible higher order correlators. 

  In this description, the standard treatment of quantum perturbations on a classical cosmological background represents an approximation, 
  in which the true quantum state is split in a coherent part describing the background and a perturbation.
   However, this approximation is expected to ``wear out'' in time
   due to loss of coherence. For certain backgrounds this 
   can lead to a complete quantum-break \cite{Dvali:2013eja,Dvali:2014gua,Berezhiani:2016grw,Dvali:2017eba,Berezhiani:2021zst}. 
   
    In the standard background field treatment, in principle, this breakdown can 
    be captured by the back-reaction on a would-be classical background 
    and should manifest itself in the relative growth of higher order 
    correlators \cite{Berezhiani:2021gph}. 

 However, we would like to separate the issues of break-down of 
 perturbative treatment on large time scales from the 
 questions of differences in gauge-fixings in the two approaches.    
 In our approach the procedure must begin with the gauge-fixing sector in place before introducing the nontrivial background state for the degrees of freedom in question.
 
 Next step is the construction of the cosmological 
 background state. This requires a source. The role 
 of the source can be played by an additional massive scalar 
field with the following Lagrangian:  
\beq
\Delta \mathcal{L}=\sqrt{-g}\left( -\frac{1}{2}g^{\mu\nu}\partial_\mu\phi \partial_\nu\phi-\frac{1}{2}m^2\phi^2\right)\,.
\eeq
Upon canonical quantization, the vacuum of the theory would still correspond to Minkowski spacetime with properties
\beq
\la\Omega | \hat{g}\mn |\Omega\ra=\eta\mn\,,\qquad \la\Omega | \hat{\phi} |\Omega\ra=0\,.
\eeq
However, there should also exist states in the physical Hilbert space for which $\la  \hat{\phi} \ra\neq 0$. And such states are expected to source the non-trivial state of gravitational degrees of freedom with $\la \hat{g}\mn \ra\neq\eta\mn$.

Let us consider the state that classically corresponds to spatially flat cosmology with the spacetime interval
\beq
\label{FRW}
ds^2=-dt^2+a^2(t)dx^2\,,
\eeq
with $a(t)$ being the scale factor; e.g., during quasi-de Sitter evolution, like cosmic inflation, $a(t)\simeq e^{Ht}$ with high accuracy.

In quantum theory, on the other hand, we are interested in quantum (presumably coherent) states, in which the expectation value of the metric operator $\la \hat{g}_{\mu\nu}\ra$ reproduces the desired classical geometry. Again, it is important to appreciate that one has to quantize the theory before constructing the above state. 

The BRST quantization begins with the introduction of the gauge-fixing sector. We begin by considering the de Donder gauge. 
Let us assume that there exists a BRST-invariant state $|g_{\rm cl}\ra$, such that $\la g_{\rm cl}|\hat{g}_{\mu\nu}|g_{\rm cl}\ra=g^{\rm cl}_{\mu\nu}$ reproduces the aforementioned FLRW metric in the classical limit. As we have already argued above, de Donder gauge 
implies \eqref{dedondercl} up to $\mathcal{O}(\hbar)$-corrections. It is straightforward to see that this condition is not satisfied by \eqref{FRW}, implying that we will be deviating from the corresponding slicing dynamically. If we would like to maintain classical FLRW background geometry, an appropriate gauge-fixing condition is required. For example, the choice we have considered for the Rindler background \eqref{cosmogauge} also works for FLRW slicing \eqref{FRW}.

 With the proper gauge-fixing at hand, we can proceed with 
 computations of correlators order by order in 
 $\hbar$-expansion. This procedure must correctly match 
 the standard background field method for accounting for  
 the quantum back-reaction in the specific gauge \eqref{cosmogauge}.

\subsection{A Class of Pure-Gauge States}

Although, in general, a construction of BRST-invariant states 
in quantum gravity can be highly involved, for illustrative purposes 
we shall produce a simple class of manifestly BRST-invariant states.
These are the states built by $\hat{b}_\nu$ which are invariant under BRST transformations, in complete analogy with QED \cite{BDS_QED}. Despite the modification of 1-point expectation value and correspondingly the background, the utilization of $\hat{b}_\nu$ is incapable of generating a physically distinct states because of \eqref{bQc}.

Let us consider the following coherent state
\beq
\label{pure_gauge_gr}
|f\ra=e^{-i\int d^3 x f^\nu_c(x)\hat{b}_\nu}|\Omega\ra\,,
\eeq
where $f^\nu_c(x)$ are the c-number functions of spacial coordinates, while $|\Omega\ra$ stands for the Minkowski vacuum; i.e. $\la \Omega | \hat{g}_{\mu\nu} |\Omega\ra=\eta_{\mu\nu}$. Despite similarities, it must be mentioned that this state differs from a similar one constructed in QED. Namely, $\hat{b}_\nu$ is not {\it per se} a conjugate momentum for the temporal degree of freedom $\hat{A}^\nu$, because of the participation of ghosts in \eqref{pinu}. However, it is easy to see that this difference does not generate a ghost number, nor any other nontrivial ghost configuration.

It is a direct consequence of \eqref{bQc} and the nilpotence of the BRST charge ($\hat{Q}^2=0$) that \eqref{pure_gauge_gr} is physically equivalent to the Minkowski vacuum, i.e.
\beq
|f\ra=|\Omega\ra+\hat{Q}|\psi\ra\,,
\eeq
with some, not necessarily a physical, state $|\psi\ra$.
As usual, this implies that this state has the same $S$-matrix elements with physical states as $|\Omega\ra$. Similarly we could replace $|\Omega\ra$ in \eqref{pure_gauge_gr} with any other physical state and the result of the exponential operator would be equivalent.

Yet, the expectation value of the metric is modified from its Minkowskian value
\beq
\la f |A^\mu| f\ra=-\delta^{\mu}_0+f^\mu_c \,,\qquad \la f |g_{ij}| f\ra=\delta_{ij}\,.
\eeq
The above indicates that the coherent state in question corresponds to a Minkowski space in different coordinates. 

It is important to appreciate that this class of pure gauge states is quite limited and by no means guarantees that a particular slicing of the Minkowski spacetime can be realized within it. For example, it may naively seem that the quantum state corresponding to the Rindle space \eqref{rindler} may be consistently captured within \eqref{pure_gauge_gr}. However, the coordinate singularity of that frame complicates the construction. Recalling that in de Donder gauge $A^\mu=\sqrt{-g}g^{0\mu}$, we can easily find $f_c^\mu$ that would correspond to \eqref{rindler}. In this gauge, however, the metric would depart from \eqref{rindler} classically at later times.

Therefore, in order  for the classical dynamics not to take us away from the Rindler coordinates, it is better to work with \eqref{cosmogauge}, in which case $A^\mu=g^{0\mu}$. The resulting coherent state corresponding to \eqref{rindler} would be
\beq
|R\ra=e^{-i\int d^3x\left[1+(ax)^{-2}\right]\hat{b}_0}|\Omega\ra\,.
\eeq
Notice that the state becomes singular at the horizon, which invalidates the construction. In other words, it does not seem possible to represent \eqref{rindler} as a coherent state (of the form \eqref{pure_gauge_gr}) built over the Minkowski space. 

\section{On Cutoff-Sensitivity and Supersymmetry}
\label{cutoff}

  We would like to comment on the issue of cutoff-sensitivity of 
  the low energy theory.   First of all, let us remark that the presented quantization on Minkowski 
  vacuum can be performed without any reference to the cutoff,  
  the value of which depends on the details of the theory. 
 While the Planck mass sets an upper bound on this scale,  
with the extended field content the cutoff can be much lower. 
 In general,  the cutoff of a low energy effective theory propagating 
 $N_{\rm sp}$ particle species, is lowered by the black hole 
 physics to the ``species scale'',  $\mpl/\sqrt{N_{\rm sp}}$
 \cite{Dvali:2007hz, Dvali:2007wp, Dvali:2008ec, Dvali:2009ks, Dvali:2010vm}.

  Although our quantization is not sensitive to the cutoff, upon 
  computation of higher order corrections, one inevitably encounters 
  such questions. 
   However, the contributions that are cutoff-sensitive must be
   reliably evaluated within the UV-complete theory.
  
At the level of the low energy effective theory the 
integration-out of UV-physics  
must result into an infinite series of operators suppressed by
the cutoff.  Their effects on the low energy observables, 
such as, e.g., planetary orbits, depend on the scale of the 
problem (e.g., size of the orbit) relative the cutoff length and is a matter of phenomenology. 
   For reasonable values of the cutoff scale, for most of the macroscopic systems of astrophysical interest such corrections are negligible. 
    In this sense, the effects of UV-completion on the low energy 
    observables are sub-dominant relative to Einstein.

  However, interestingly, our framework imposes non-trivial consistency constraints on certain would-be cutoff-sensitive parameters. 
   An important example is provided by the 
  vacuum energy, $\Lambda$.  At the level of loop-expansion,
  the contributions must come from the  
  diagrams with external legs that in the classical effective 
  action get re-summed into 
  $\sqrt{-g} \Lambda$ term.  This is imposed by the general covariance of the  effective action obtained in the classical limit. 
  
   Again, due to cutoff-sensitivity the computation must be performed 
   within an UV-complete theory.  However, even without knowing this theory,  in our framework $\Lambda$ is nullified by consistency. 
    This is because the quantization on top of exact Minkowski vacuum is our defining point.   A would-be shift of this vacuum to 
    either de Sitter or anti-de Sitter (AdS), would render our scheme self-contradictory. 
Thereby, $\Lambda =0$ must be the exact non-perturbative quantum 
 vacuum in our theory. 

 The above conclusion is identical to the condition  $\Lambda =0$ imposed  by the requirement 
 of the consistent $S$-matrix (see \cite{Dvali:2020etd} and references therein). This is not a coincidence
 as the Minkowski vacuum is required both as the $S$-matrix vacuum
that admits a globally-defined time, as well as, the 
vacuum for consistent quantization of asymptotic $S$-matrix 
states.  
 
 Notice that, although none of the above constraints exclude 
 the existence of AdS states, they do exclude the transitions 
 into such states from our Minkowski vacuum. 
 In other words, the AdS vacua permitted by our quantization 
 must not jeopardize the stability of the Minkowski one. 
 In this sense, the two must belong to different superselection sectors. 

  We remark that impossibility of decay of the Minkowski vacuum into 
  an AdS one is also supported by the $S$-matrix considerations 
  as it is impossible to consistently describe such a decay 
  as a unitary $S$-matrix process \cite{Dvali:2011wk, Zeldovich:1974py}. 

  The prominent role of the Minkowski vacuum is intrinsically intertwined 
 with the question of supersymmetry.  First, the key feature of the valid Minkowski vacuum is its stability. This can be guaranteed by the Poincar\'e invariance. Notice that this requirement is exact and must go beyond the perturbative level.  That is, the decay of the Minkowski vacuum via non-perturbative tunneling  processes is also excluded. 
   In particular, this implies \cite{Dvali:2011wk} that a consistent 
 theory must exclude the existence of a neighboring AdS vacuum 
 with the energy splitting exceeding the Coleman-De Luccia \cite{Coleman:1980aw} tunneling threshold.
 The absence of tunneling gives a first indication of the potential role 
 of supersymmetry. 
Namely, supersymmetry plays a prominent role in guaranteeing the exact Poincar\'e invariance of the Minkowski vacuum.  

 However, there is more to it, as supersymmetry appears to be a 
 necessity rather than merely a stabilizing tool. 
 Even in the absence of any scalar potential or a cosmological term at the perturbative level, supersymmetry appears to be necessary for excluding the violation of Poincar\'e symmetry by the gravitational instantons \cite{Dvali:2024dlb}. As shown in the latter work, even in pure gravity, the Poincar\'e symmetry of Minkowski vacuum is inevitably violated (by generating a continuum of the eternal de Sitter states)  by Eguchi-Hanson instantons \cite{Eguchi:1978xp,Eguchi:1978gw} which are the intrinsic feature of 
 the topological structure of such a vacuum.  Supersymmetry
 restores the Poincar\'e symmetry thanks to the 
 fermionic zero modes deposited in Eguchi-Hanson instantons by the spin-$3/2$ gravitino. Simultaneously, gravitino gets a mass, thereby 
 signaling the super-Higgs effect. 
 
In the light of the above arguments, our quantization framework provides a strong motivation for supersymmetry.

\section{Conclusion}

\label{conclusion}

In this work we have performed the canonical quantization of GR in the BRST-invariant framework.  We demonstrated the recovery of dynamical classical properties of GR in the appropriate limit. 
This is of central importance, since a quantization of 
GR within effective field theory is usually believed to be problematic because of the constraint structure.

We showed that the BRST-invariant approach \cite{Kugo:1978rj} is fully consistent with the existence of the Hamiltonian time-flow in the physical Hilbert space. We pinpoint the place where the over-constraining of the physical states takes place in other approaches and how they differ from the well-known, and well-tested, procedures of handling constraints in gauge theories.

    Our framework sheds a very different light on the role of 
    the background in quantum gravity. 
The distinguishing feature is that the theory is quantized with the gauge-fixing in place before we select a quantum state that corresponds to a  nontrivial (non-Minkowski) spacetime. 
 This solidifies the idea that in 
 quantum gravity the curved classical backgrounds must be treated 
 as coherent states constructed on top of the Minkowski vacuum 
\cite{Dvali:2011aa,Dvali:2013eja,Dvali:2014gua,Berezhiani:2016grw,Dvali:2017eba,Dvali:2020etd,Berezhiani:2021zst}. In particular, the BRST 
invariant construction of a de Sitter state can be found in 
\cite{Berezhiani:2021zst}.

 Naturally, the Minkowski spacetime is given a well-deserved special treatment as the proclaimed vacuum of quantum gravity. 
 In this picture the characteristic ``democracy'' of 
GR, according to which any consistent classical 
metric can be treated as a valid vacuum for quantum perturbations, 
is the emergent property of coherent states in the proper limit.
 In this way, the fundamental gauge-fixing term is universally 
 consistent for all physical spacetimes.

This picture has number of implications, in particular for cosmology. 
The coherent state description of a cosmological space time in full quantum theory can uncover new properties that are invisible in the standard semi-classical treatment. The example is provided by quantum break-time of de Sitter,  which shows that the eternal de Sitter is not a valid vacuum of quantum gravity \cite{Dvali:2013eja,Dvali:2014gua,Berezhiani:2016grw,Dvali:2017eba,
Dvali:2018ytn, Dvali:2021bsy,
Berezhiani:2021zst}.

 The prominent status of Minkowski as of the vacuum resonates 
 with formulation of quantum gravity via an $S$-matrix theory
 in which in and out states are given by a complete set of asymptotic states defined on Minkowski. 

 Although, the presented framework is independent of the 
 UV-completion of the theory,  
 the above-discussed conceptual features of our  
 framework are validated by the string theory in very general terms.  
 
  In particular, the string theory makes manifest both the defining role of the $S$-matrix and the coherent state interpretation of cosmological backgrounds \cite{Dvali:2020etd}. For example, the de Sitter-like inflationary backgrounds 
  in string theory are obtained via $D$-branes \cite{Dvali:1998pa, Dvali:2001fw}, which are the only known string-theoretic sources with negative pressure. However,  $D$-branes \cite{Polchinski:1995mt} represent the string solitons and thus the coherent states of stringy degrees of freedom.
  This makes the coherent state view of the inflationary cosmological background \cite{Dvali:2013eja,Dvali:2014gua,Berezhiani:2016grw,Dvali:2017eba,
Dvali:2018ytn, Dvali:2021bsy,
Berezhiani:2021zst} organic for string theory.

  The last but not least, our quantization framework, due 
to the crucial  role of the Minkowaski vacuum, provides strong motivation 
for supersymmetry.  In particular,  as shown in \cite{Dvali:2024dlb},
without supersymmetry, the Poincar\'e invariance would be inevitably broken at a non-perturbative level by the Eguchi-Hanson instantons (due to generation of the de Sitter ``vacua''). 
These instantons are fundamentally linked with the topological structure 
of the Minkowski vacuum in gravity and their existence is insensitive 
to UV-properties of the theory.  Therefore, they must be removed within 
low energy effective field theory.  This is accomplished by the 
fermionic zero-modes of spin-$3/2$ gravitino, thereby 
justifying supersymmetry.  It is important to stress that 
supersymmetry  must be spontaneously broken by consistency of the instanton-generated gravitino mass with the Poincar\'e symmetry. 
 That is, the low energy effective theory 
 below gravitino mass is Einstein's gravity with an 
 exact Poincar\'e-invariant vacuum.

 Supersymmetry is also instrumental for eliminating other processes of 
 violation of the Poincar\'e symmetry, such as via tunneling, which 
 is essential for having a 
 consistent theory with Minkowski vacuum \cite{Dvali:2011wk, Zeldovich:1974py}. 

  Thus, via the requirement of exactness of the Poincar\'e symmetry of the Minkowski vacuum, our framework supports a spontaneously-broken supersymmetry.

\section*{Acknowledgements}

We would like to thank Giordano Cintia, Giacomo Contri and Archil Kobakhidze for stimulating and fruitful discussions. O.S. would like to thank the Arnold Sommerfeld Center at LMU and the Max-Planck-Institute for Physics, Munich for the hospitality extended during the completion of this work. This work was supported in part by the Humboldt Foundation under the Humboldt Professorship Award, by the European Research Council Gravities Horizon Grant AO number: 850 173-6, by the Deutsche Forschungsgemeinschaft (DFG, German Research Foundation) under Germany’s Excellence Strategy - EXC-2111 - 390814868, Germany’s Excellence Strategy under Excellence Cluster Origins  EXC 2094 – 390783311 and 
the Australian Research Council under the Discovery Projects grants DP210101636 and DP220101721.     

\hskip 10pt

\noindent {\bf Disclaimer:} Funded by the European Union. Views and opinions expressed are however those of the authors only and do not necessarily reflect those of the European Union or European Research Council. Neither the European Union nor the granting authority can be held responsible for them.


\begin{thebibliography}{99}


\bibitem{Dvali:2011aa}
G.~Dvali and C.~Gomez,
``Black Hole's Quantum N-Portrait,''
Fortsch. Phys. \textbf{61}, 742-767 (2013)
doi:10.1002/prop.201300001
[arXiv:1112.3359 [hep-th]].


\bibitem{Dvali:2012en}
G.~Dvali and C.~Gomez,
``Black Holes as Critical Point of Quantum Phase Transition,''
Eur. Phys. J. C \textbf{74}, 2752 (2014)
doi:10.1140/epjc/s10052-014-2752-3
[arXiv:1207.4059 [hep-th]].


\bibitem{Dvali:2013eja}
G.~Dvali and C.~Gomez,
``Quantum Compositeness of Gravity: Black Holes, AdS and Inflation,''
JCAP \textbf{01}, 023 (2014)
doi:10.1088/1475-7516/2014/01/023
[arXiv:1312.4795 [hep-th]].


\bibitem{Dvali:2014gua}
G.~Dvali and C.~Gomez,
``Quantum Exclusion of Positive Cosmological Constant?,''
Annalen Phys. \textbf{528}, 68-73 (2016)
doi:10.1002/andp.201500216
[arXiv:1412.8077 [hep-th]].


\bibitem{Berezhiani:2016grw}
L.~Berezhiani,
``On Corpuscular Theory of Inflation,''
Eur. Phys. J. C \textbf{77}, no.2, 106 (2017)
doi:10.1140/epjc/s10052-017-4672-5
[arXiv:1610.08433 [hep-th]].


\bibitem{Dvali:2017eba}
G.~Dvali, C.~Gomez and S.~Zell,
``Quantum Break-Time of de Sitter,''
JCAP \textbf{06}, 028 (2017)
doi:10.1088/1475-7516/2017/06/028
[arXiv:1701.08776 [hep-th]].



\bibitem{Berezhiani:2021zst}
L.~Berezhiani, G.~Dvali and O.~Sakhelashvili,
``de Sitter space as a BRST-invariant coherent state of gravitons,''
Phys. Rev. D \textbf{105}, no.2, 025022 (2022)
doi:10.1103/PhysRevD.105.025022
[arXiv:2111.12022 [hep-th]].


\bibitem{Dvali:2017ruz}
G.~Dvali and S.~Zell,
``Classicality and Quantum Break-Time for Cosmic Axions,''
JCAP \textbf{07}, 064 (2018)
doi:10.1088/1475-7516/2018/07/064
[arXiv:1710.00835 [hep-ph]].


\bibitem{Dvali:2013vxa}
G.~Dvali, D.~Flassig, C.~Gomez, A.~Pritzel and N.~Wintergerst,
``Scrambling in the Black Hole Portrait,''
Phys. Rev. D \textbf{88}, no.12, 124041 (2013)
doi:10.1103/PhysRevD.88.124041
[arXiv:1307.3458 [hep-th]].


\bibitem{Berezhiani:2020pbv}
L.~Berezhiani and M.~Zantedeschi,
``Evolution of coherent states as quantum counterpart of classical dynamics,''
Phys. Rev. D \textbf{104}, no.8, 085007 (2021)
doi:10.1103/PhysRevD.104.085007
[arXiv:2011.11229 [hep-th]].


\bibitem{Berezhiani:2021gph}
L.~Berezhiani, G.~Cintia and M.~Zantedeschi,
``Background-field method and initial-time singularity for coherent states,''
Phys. Rev. D \textbf{105}, no.4, 045003 (2022)
doi:10.1103/PhysRevD.105.045003
[arXiv:2108.13235 [hep-th]].


\bibitem{Dvali:2022vzz}
G.~Dvali and L.~Eisemann,
``Perturbative understanding of nonperturbative processes and quantumization versus classicalization,''
Phys. Rev. D \textbf{106}, no.12, 125019 (2022)
doi:10.1103/PhysRevD.106.125019
[arXiv:2211.02618 [hep-th]].


\bibitem{Berezhiani:2023uwt}
L.~Berezhiani, G.~Cintia and M.~Zantedeschi,
``Perturbative construction of coherent states,''
Phys. Rev. D \textbf{109}, no.8, 085018 (2024)
doi:10.1103/PhysRevD.109.085018
[arXiv:2311.18650 [hep-th]].


\bibitem{Birrell:1982ix}
N.~D.~Birrell and P.~C.~W.~Davies,
``Quantum Fields in Curved Space,''
Cambridge Univ. Press, 1984,
ISBN 978-0-521-27858-4, 978-0-521-27858-4
doi:10.1017/CBO9780511622632


\bibitem{Dvali:2018ytn}
G.~Dvali, L.~Eisemann, M.~Michel and S.~Zell,
``Universe's Primordial Quantum Memories,''
JCAP \textbf{03}, 010 (2019)
doi:10.1088/1475-7516/2019/03/010
[arXiv:1812.08749 [hep-th]].


\bibitem{Dvali:2018xpy}
G.~Dvali,
``A Microscopic Model of Holography: Survival by the Burden of Memory,''
[arXiv:1810.02336 [hep-th]].


\bibitem{Dvali:2020wft}
G.~Dvali, L.~Eisemann, M.~Michel and S.~Zell,
``Black hole metamorphosis and stabilization by memory burden,''
Phys. Rev. D \textbf{102}, no.10, 103523 (2020)
doi:10.1103/PhysRevD.102.103523
[arXiv:2006.00011 [hep-th]].


\bibitem{Dvali:2024hsb}
G.~Dvali, J.~S.~Valbuena-Berm\'udez and M.~Zantedeschi,
``Memory burden effect in black holes and solitons: Implications for PBH,''
Phys. Rev. D \textbf{110} (2024) no.5, 056029
doi:10.1103/PhysRevD.110.056029
[arXiv:2405.13117 [hep-th]].


\bibitem{Alexandre:2024nuo}
A.~Alexandre, G.~Dvali and E.~Koutsangelas,
``New mass window for primordial black holes as dark matter from the memory burden effect,''
Phys. Rev. D \textbf{110}, no.3, 036004 (2024)
doi:10.1103/PhysRevD.110.036004
[arXiv:2402.14069 [hep-ph]].


\bibitem{Thoss:2024hsr}
V.~Thoss, A.~Burkert and K.~Kohri,
``Breakdown of hawking evaporation opens new mass window for primordial black holes as dark matter candidate,''
Mon. Not. Roy. Astron. Soc. \textbf{532}, no.1, 451-459 (2024)
doi:10.1093/mnras/stae1098
[arXiv:2402.17823 [astro-ph.CO]].


\bibitem{Balaji:2024hpu}
S.~Balaji, G.~Dom\`enech, G.~Franciolini, A.~Ganz and J.~Tr\"ankle,
``Probing modified Hawking evaporation with gravitational waves from the primordial black hole dominated universe,''
[arXiv:2403.14309 [gr-qc]].


\bibitem{Haque:2024eyh}
M.~R.~Haque, S.~Maity, D.~Maity and Y.~Mambrini,
``Quantum effects on the evaporation of PBHs: contributions to dark matter,''
JCAP \textbf{07}, 002 (2024)
doi:10.1088/1475-7516/2024/07/002
[arXiv:2404.16815 [hep-ph]].


\bibitem{Berezhiani:2015ola}
L.~Berezhiani and M.~Trodden,
``How Likely are Constituent Quanta to Initiate Inflation?,''
Phys. Lett. B \textbf{749}, 425-430 (2015)
doi:10.1016/j.physletb.2015.08.007
[arXiv:1504.01730 [hep-th]].


\bibitem{Berezhiani:2022gnv}
L.~Berezhiani and M.~Trodden,
``A relativistic gas of inflatons as an initial state for inflation,''
Phys. Lett. B \textbf{840}, 137852 (2023)
doi:10.1016/j.physletb.2023.137852
[arXiv:2211.06222 [hep-th]].


\bibitem{Garriga:2007zk}
J.~Garriga and T.~Tanaka,
``Can infrared gravitons screen Lambda?,''
Phys. Rev. D \textbf{77}, 024021 (2008)
doi:10.1103/PhysRevD.77.024021
[arXiv:0706.0295 [hep-th]].


\bibitem{Tsamis:1996qq}
N.~C.~Tsamis and R.~P.~Woodard,
``Quantum gravity slows inflation,''
Nucl. Phys. B \textbf{474}, 235-248 (1996)
doi:10.1016/0550-3213(96)00246-5
[arXiv:hep-ph/9602315 [hep-ph]].


\bibitem{Tsamis:1996qm}
N.~C.~Tsamis and R.~P.~Woodard,
``The Quantum gravitational back reaction on inflation,''
Annals Phys. \textbf{253}, 1-54 (1997)
doi:10.1006/aphy.1997.5613
[arXiv:hep-ph/9602316 [hep-ph]].


\bibitem{Woodard:2004ut}
R.~P.~Woodard,
``de Sitter breaking in field theory,''
[arXiv:gr-qc/0408002 [gr-qc]].


\bibitem{Tsamis:2007is}
N.~C.~Tsamis and R.~P.~Woodard,
``Comment on `Can infrared gravitons screen Lambda?',''
Phys. Rev. D \textbf{78}, 028501 (2008)
doi:10.1103/PhysRevD.78.028501
[arXiv:0708.2004 [hep-th]].


\bibitem{Weinberg}
S.~Weinberg, ``The quantum theory of fields. Vol. 2: Modern applications'', Cambridge, UK: University Press (1996)



\bibitem{Dvali:2020etd}
G.~Dvali,
``$S$-Matrix and Anomaly of de Sitter,''
Symmetry \textbf{13}, no.1, 3 (2020)
doi:10.3390/sym13010003
[arXiv:2012.02133 [hep-th]].


\bibitem{Dvali:2024dlb}
G.~Dvali, A.~Kobakhidze and O.~Sakhelashvili,
``Hint to Supersymmetry from GR Vacuum,''
[arXiv:2406.18402 [hep-th]].


\bibitem{Veltman:1975vx}
M.~J.~G.~Veltman,
``Quantum Theory of Gravitation,''
Conf. Proc. C \textbf{7507281}, 265-327 (1975)


\bibitem{Kugo:1978rj}
T.~Kugo and I.~Ojima,
``Subsidiary Conditions and Physical S Matrix Unitarity in Indefinite Metric Quantum Gravitational Theory,''
Nucl. Phys. B \textbf{144}, 234-252 (1978)
doi:10.1016/0550-3213(78)90504-7


\bibitem{Dvali:lecture}
G.~Dvali, ``Black Holes from Particle Physics Perspective'', 2014, https://cds.cern.ch/record/1970923



\bibitem{Deffayet:2001uk}
C.~Deffayet, G.~R.~Dvali, G.~Gabadadze and A.~I.~Vainshtein,
``Nonperturbative continuity in graviton mass versus perturbative discontinuity,''
Phys. Rev. D \textbf{65}, 044026 (2002)
doi:10.1103/PhysRevD.65.044026
[arXiv:hep-th/0106001 [hep-th]].





\bibitem{Kugo:1979gm}
T.~Kugo and I.~Ojima,
``Local Covariant Operator Formalism of Nonabelian Gauge Theories and Quark Confinement Problem,''
Prog. Theor. Phys. Suppl. \textbf{66}, 1-130 (1979)
doi:10.1143/PTPS.66.1


\bibitem{VanHolten:2001nj}
J.~W.~Van Holten,
``Aspects of BRST quantization,''
Lect. Notes Phys. \textbf{659}, 99-166 (2005)
doi:10.1007/978-3-540-31532-2\_3
[arXiv:hep-th/0201124 [hep-th]].


\bibitem{Witten:2022xxp}
E.~Witten,
``A Note On The Canonical Formalism for Gravity,''
[arXiv:2212.08270 [hep-th]].


\bibitem{Arnowitt:1959ah}
R.~L.~Arnowitt, S.~Deser and C.~W.~Misner,
``Dynamical Structure and Definition of Energy in GR,''
Phys. Rev. \textbf{116}, 1322-1330 (1959)
doi:10.1103/PhysRev.116.1322


\bibitem{Witten:1981mf}
E.~Witten,
``A Simple Proof of the Positive Energy Theorem,''
Commun. Math. Phys. \textbf{80} (1981), 381
doi:10.1007/BF01208277


\bibitem{DeWitt:1967yk}
B.~S.~DeWitt,
``Quantum Theory of Gravity. 1. The Canonical Theory,''
Phys. Rev. \textbf{160}, 1113-1148 (1967)
doi:10.1103/PhysRev.160.1113


\bibitem{BDS_QED}
L.~Berezhiani, G.~Dvali and O.~Sakhelashvili, in preparation.


\bibitem{Dvali:2007hz}
G.~Dvali,
``Black Holes and Large N Species Solution to the Hierarchy Problem,''
Fortsch. Phys. \textbf{58}, 528-536 (2010)
doi:10.1002/prop.201000009
[arXiv:0706.2050 [hep-th]].

 
\bibitem{Dvali:2007wp}
G.~Dvali and M.~Redi,
``Black Hole Bound on the Number of Species and Quantum Gravity at LHC,''
Phys. Rev. D \textbf{77}, 045027 (2008)
doi:10.1103/PhysRevD.77.045027
[arXiv:0710.4344 [hep-th]].



\bibitem{Dvali:2008ec}
G.~Dvali and C.~Gomez,
``Quantum Information and Gravity Cutoff in Theories with Species,''
Phys. Lett. B \textbf{674}, 303-307 (2009)
doi:10.1016/j.physletb.2009.03.024
[arXiv:0812.1940 [hep-th]].



\bibitem{Dvali:2009ks}
G.~Dvali and D.~Lust,
``Evaporation of Microscopic Black Holes in String Theory and the Bound on Species,''
Fortsch. Phys. \textbf{58}, 505-527 (2010)
doi:10.1002/prop.201000008
[arXiv:0912.3167 [hep-th]].



\bibitem{Dvali:2010vm}
G.~Dvali and C.~Gomez,
``Species and Strings,''
[arXiv:1004.3744 [hep-th]].


\bibitem{Dvali:2011wk}
G.~Dvali,
``Safety of Minkowski Vacuum,''
[arXiv:1107.0956 [hep-th]].


\bibitem{Zeldovich:1974py}
Y.~B.~Zeldovich,
``Spontaneous processing in vacuum,''
Phys. Lett. B \textbf{52} (1974), 341-343
doi:10.1016/0370-2693(74)90057-4


\bibitem{Coleman:1980aw}
S.~R.~Coleman and F.~De Luccia,
``Gravitational Effects on and of Vacuum Decay,''
Phys. Rev. D \textbf{21} (1980), 3305
doi:10.1103/PhysRevD.21.3305


\bibitem{Eguchi:1978xp}
T.~Eguchi and A.~J.~Hanson,
``Asymptotically Flat Selfdual Solutions to Euclidean Gravity,''
Phys. Lett. B \textbf{74} (1978), 249-251
doi:10.1016/0370-2693(78)90566-X


\bibitem{Eguchi:1978gw}
T.~Eguchi and A.~J.~Hanson,
``Selfdual Solutions to Euclidean Gravity,''
Annals Phys. \textbf{120} (1979), 82
doi:10.1016/0003-4916(79)90282-3


\bibitem{Dvali:2021bsy}
G.~Dvali,
``Quantum Gravity in Species Regime,''
[arXiv:2103.15668 [hep-th]].


\bibitem{Dvali:1998pa}
G.~R.~Dvali and S.~H.~H.~Tye,
``Brane inflation,''
Phys. Lett. B \textbf{450}, 72-82 (1999)
doi:10.1016/S0370-2693(99)00132-X
[arXiv:hep-ph/9812483 [hep-ph]].


\bibitem{Dvali:2001fw}
G.~R.~Dvali, Q.~Shafi and S.~Solganik,
``D-brane inflation,''
[arXiv:hep-th/0105203 [hep-th]].


\bibitem{Polchinski:1995mt}
J.~Polchinski,
``Dirichlet Branes and Ramond-Ramond charges,''
Phys. Rev. Lett. \textbf{75}, 4724-4727 (1995)
doi:10.1103/PhysRevLett.75.4724
[arXiv:hep-th/9510017 [hep-th]].







\end{thebibliography}
\end{document}